\documentclass[11pt]{article}

\usepackage{bm}
\usepackage{amssymb}
\usepackage{amsmath}
\usepackage{amsthm}
\usepackage{float}
\usepackage{multirow}
\usepackage{color}
\usepackage{comment}
\usepackage{fullpage}
\usepackage[normalem]{ulem} 
\usepackage{makeidx}
\usepackage{xspace}
\usepackage[bottom]{footmisc}

\makeindex

\usepackage[utf8]{inputenc}
\usepackage[margin=3cm]{geometry}
\usepackage{natbib}
\usepackage{url}
\usepackage{setspace}
\usepackage{array}
\usepackage[final]{hyperref}

\usepackage{enumitem}
\usepackage{booktabs}

\usepackage{graphicx}
\usepackage{caption}
\usepackage{subcaption}
\usepackage[ruled,vlined]{algorithm2e}
\usepackage{longtable}

\allowdisplaybreaks

\setcitestyle{authoryear,open={(},close={)}}

\title{Shrinkage Bayesian Causal Forests for Heterogeneous Treatment Effects Estimation \thanks{Earlier draft of this paper was titled ``Sparse Bayesian Causal Forests for Heterogeneous Treatment Effects Estimation".}}

\author{%
	Alberto Caron \thanks{This work was supported by a British Heart 		Foundation-Turing Cardiovascular Data Science Award (BCDSA/100003). 
		Corresponding author: \texttt{alberto.caron.19@ucl.ac.uk}, 1-19 Torrington Pl, London WC1E 7HB. } \\
	\footnotesize Department of Statistical Science\\
	\footnotesize University College London\\
	\and
	Gianluca Baio \\
	\footnotesize Department of Statistical Science\\
	\footnotesize University College London\\
	\and
	Ioanna Manolopoulou \\
	\footnotesize Department of Statistical Science\\
	\footnotesize University College London\\
}

\begin{document}
	
	\maketitle
	
	\begin{abstract}
		This paper develops a sparsity-inducing version of Bayesian Causal Forests, a recently proposed nonparametric causal regression model that employs Bayesian Additive Regression Trees and is specifically designed to estimate heterogeneous treatment effects using observational data. The sparsity-inducing component we introduce is motivated by empirical studies where not all the available covariates are relevant, leading to different degrees of sparsity underlying the surfaces of interest in the estimation of individual treatment effects. The extended version presented in this work, which we name Shrinkage Bayesian Causal Forest, is equipped with an additional pair of priors allowing the model to adjust the weight of each covariate through the corresponding number of splits in the tree ensemble. These priors improve the model's adaptability to sparse data generating processes and allow to perform fully Bayesian feature shrinkage in a framework for treatment effects estimation, and thus to uncover the moderating factors driving heterogeneity. In addition, the method allows prior knowledge about the relevant confounding covariates and the relative magnitude of their impact on the outcome to be incorporated in the model. We illustrate the performance of our method in simulated studies, in comparison to Bayesian Causal Forest and other state-of-the-art models, to demonstrate how it scales up with an increasing number of covariates and how it handles strongly confounded scenarios. Finally, we also provide an example of application using real-world data. 
	\end{abstract}

	\noindent%
	\vfill
	
	\newpage
	
	\section{Introduction}
	
	Inferring the treatment effect at an individual level in a population of interest lies at the heart of disciplines such as precision medicine or personalized advertisement, where decision making in terms of treatment administration is based on individual characteristics. The ever-increasing amount of observational data available offers a unique opportunity for drawing inferences at the resolution of each individual. However, since Individual Treatment Effects (ITEs) are never directly observable in the real world, standard supervised learning techniques cannot be directly applied. Moreover, the process of treatment allocation in large, observational datasets is usually unknown and can obscure the effect of the actual treatment through confounding \citep{dawid_2000, pearl_2009, imbens_rubin_2015}.

	The application of statistical learning tools for causal inference has led to significant improvements in the estimation of heterogeneous treatment effects. These improvements stem from the predictive power of advanced nonparametric regression models, that, after being appropriately adapted to the causal inference setting, can leverage large observational datasets and capture non-linear relationships. \cite{caron_2020} provide a review of the most recent and popular methods, together with a comparison of their performance.
	
	Two of the early contributions that paved the way towards the use of tree-based statistical learning tools on large datasets for causal analysis purposes are \cite{foster_2011} and \cite{hill_2011}, who advocate the use of tree ensemble methods for the estimation of ITE. The former focuses on randomized experiments and makes use of Random Forests \citep{breiman_2001}; the latter instead addresses the problem from an observational study perspective, and employs Bayesian Additive Regression Trees (BART) \citep{cgm_1998, chipman_2010}. Another early contribution that focuses instead on Average Treatment Effect (ATE) estimation in observational studies is \cite{traskin_2011}, where the authors propose classification trees for identifying the study population with sufficient overlap. A more recent and popular tree-based method for ITE estimation is Causal Forests (CF) \citep{athey_2019}, a causal implementation of Random Forests. \cite{hahn_2020} instead build on the work of \cite{chipman_2010} and \cite{hill_2011} to formulate a new BART framework for causal analysis, under the name of Bayesian Causal Forests (BCF), specifically designed to address strong confounding and separate between prognostic and moderating effects of the covariates when estimating ITE. The prognostic effect (or prognostic score) is defined as the impact of the covariates on the outcome in the absence of treatment, while the moderating effect is the impact of the covariates on the response to treatment.

	A different stream of contributions that do not focus on any specific regression model is that of \emph{Meta-Learners}. Meta-Learners are meta-algorithms that design a procedure to estimate ITE via any suitable off-the-shelf supervised regression model (e.g.~random forests, neural networks, etc.). Recent popular work on Meta-Learners include \cite{kunzel_2017}, where the authors develop a framework to deal with unbalanced treatment groups (named X-Learner), and \cite{wager_2019}, where parametrization in \cite{robinson_1988} is exploited (hence the name R-Learner) to design a direct loss function on the treatment effect surface for parameter tuning. Among other notable works on ITE estimation, \cite{vanderschaar_2017, vanderschaar_2018} adopt a multi-task learning approach using Gaussian Processes, while \cite{johansson_2016}, \cite{shalit_2017} and \cite{yao_2018} employ deep neural networks to learn balanced representations that aim at minimizing a distributional distance between treatment groups. Moreover, contributions such as those of \cite{zhao_2018, zimmert_2019, fan_2020} focus specifically on high-dimensional settings, where a large number of covariates is available. In particular, \cite{zhao_2018} use \cite{robinson_1988} decomposition to estimate nuisance parameters with machine learning methods and isolate the treatment effect function, which is then fit via LASSO regression, for a more interpretable output on effect modifiers. \cite{zimmert_2019} and \cite{fan_2020} instead propose and derive properties of a cross-validated two stage estimator where nuisance parameters are fitted with ML methods in the first stage, and a nonparametric local kernel smoother is instead applied to fit treatment effect.
	
	Regression-based methods for treatment effect estimation typically leverage large samples in observational studies. However, large observational data often also feature a large number of pre-treatment covariates, many of which may not affect the response variable in question nor act as a modifier of the treatment effect. Hence, the task of estimating ITE, whose complexity inevitably depends on the smoothness and sparsity of the outcome surface \citep{vanderschaar_2018}, necessitates regularization. At the same time, prior subject-matter knowledge on the relative importance of the covariates may be available, and can improve estimates if embedded in the model. In light of these considerations, none of the aforementioned approaches mentioned allow to jointly: i) account for heterogeneous smoothness and sparsity across covariates; ii) tease apart prognostic and moderating covariates through targeted feature shrinkage; iii) incorporate prior knowledge on the relevant covariates and their relative impact on the outcome. Carefully designed regularization can lead to improved ITE estimates and inferences on prognostic and moderating factors, since including a large number of covariates in a fully-saturated model to adjust for confounding may lead to misspecification. In this work, we propose an extension of the Bayesian Causal Forest framework \citep{hahn_2020}, consisting in additional Dirichlet priors placed on the trees splitting probabilities \citep{linero_2018_sparse}, that implement fully Bayesian feature shrinkage on the prognostic and moderating covariates, and allow the incorporation of prior knowledge on their relative importance. BART (and consequently BCF) was originally designed to adapt to smoothness but not to sparsity\footnote{Regularization in BART is introduced via shallow trees structures, to avoid overfitting (similarly to Gradient Boosting). \cite{linero_smooth} proposed a way to further enhance smoothness adaptation in BART through a probabilistic version of the trees, where inputs follow a probabilistic, rather than deterministic, path to the terminal nodes.}. Our extended version of the model can be easily fitted with a slight modification of the existing MCMC algorithm and provably results in improved performance thanks to its better adaptability to sparse DGPs, for negligible extra computational cost.
	
	The rest of the paper is organized as follows. Section \ref{sec:probframe} introduces the problem of estimating treatment effects using the Neyman-Rubin causal model framework, and formulates the necessary assumptions to recover treatment effect estimates under confounded observational data. Section \ref{sec:BARTCI} offers an overview on Bayesian Additive Regression Trees and their popular causal version, Bayesian Causal Forest. Section \ref{sec:sparsetree} introduces our shrinkage-inducing extension, under the name of ``Shrinkage Bayesian Causal Forest''. Section \ref{sec:simul} presents results from simulated studies carried out to compare Shrinkage Bayesian Causal Forest performance with other state-of-the-art models. Section \ref{sec:realworld} provides an example of analysis using data from the Infant Health and Development Program aimed at investigating the effects of early educational support on cognitive abilities in low birth weight infants. Section \ref{sec:conclu} concludes with a discussion.

	\section{Problem framework} \label{sec:probframe}
	
	In this section we outline the problem of deriving an estimator for ITE using observational data, utilizing the formalism of the Neyman-Rubin potential outcomes framework \citep{rubin1978, imbens_rubin_2015}\footnote{Note that identification of causal effects can be achieved also with other causal frameworks, such as \emph{do}-calculus in Structural Causal Models \citep{pearl_2009}, or decision-theoretic approach \citep{dawid_2000, dawid_2014}, and the contribution of this work, which concerns solely estimation, still apply.}. We consider a setup where the outcome variable is continuous and the treatment assignment is binary (of the type exposure versus non-exposure), but most of the notions in this section can be generalized to non-continuous responses and more than two treatment arms. For each individual $i \in \{ 1,...,N \}$, the two potential outcomes are defined as $Y_i^{(Z_i)}$, where $Z_i \in \{ 0,1 \}$ is the binary treatment assignment, with $Z_i = 1$ indicating exposure to the treatment, while $Z_i = 0$ non-exposure. We consider continuous type of outcomes such that $\big( Y_i^{(0)}, Y_i^{(1)} \big) \in \mathbb{R}^2$. Given the potential outcomes and the binary treatment assignment, ITE is defined, for each individual $i$, as the difference $Y_i^{(1)} - Y_i^{(0)}$. The fundamental problem of causal inference is that, for each $i$, we get to observe only one of the two potential outcomes $\big( Y_i^{(0)}, Y_i^{(1)} \big) \in \mathbb{R}^2$, specifically the one corresponding to the realization of $Z_i$, i.e.~$Y_i = Z_i Y_i^{(1)} + (1 - Z_i) Y_i^{(0)}$, so that ITE is never observable. 
	
	Given a dataset $ \{ \bm{X}_i, Z_i, Y_i \} $ of sample size $N$, where $\bm{X}_i \in \mathcal{X}$ are $P$ pre-treatment covariates, the ITE is the (unobserved) difference  $Y_i^{(1)} - Y_i^{(0)}$. In practice, the goal is often to estimate the \emph{Conditional Average Treatment Effects} (CATE), defined as
	\begin{equation} \label{CATE}
	\tau(\bm{x}_i) = \mathbb{E} \Big[ Y_i^{(1)} - Y_i^{(0)} \mid \bm{X}_i = \bm{x}_i \Big] ~ .
	\end{equation}
	CATE is the conditional mean of the ITE for the given value of the covariates, averaging across individual-level noise, so it is the best estimator for ITE in terms of mean squared error.
	
	In order to estimate $\tau(\bm{x}_i)$ through the observational quantities $ \{ \bm{X}_i, Z_i, Y_i \} $, we rely on a common set of assumptions to achieve identification. First of all, as already implied by the notation introduced above, we are assuming that \emph{Stable Unit Treatment Value Assumption} (SUTVA) holds, ensuring that one unit's outcome is not affected by other units' assignment to treatment (\emph{no interference}). The second assumption is unconfoundedness, which can be expressed through the conditional independence $(Y_i^{(0)}, Y_i^{(1)} ) \perp \!\!\! \perp Z_i \mid \bm{X}_i$, and it rules out the presence of unobserved common causes of $Z$ and $Y$ (i.e.~no unobserved confounders). The third and final assumption is common support, which means that all units have a probability of falling into either treatment groups which is strictly between 0 and 1. More formally, after defining the propensity score as the probability of unit $i$ being selected into treatment given $\bm{x}_i$, 
	\begin{equation}
	\pi(\bm{x}_i) = \mathbb{P} (Z_i = 1 \mid \bm{X}_i = \bm{x}_i) ~ ,
	\end{equation}
	common support implies $ \pi(\bm{x}_i) \in (0, 1)$ $\forall i \in \{ 1, ..., N \}$, so that there is no deterministic assignment to one of the groups given features $\bm{X}_i = \bm{x}_i$. Note that unconfoundedness and common support are automatically satisfied in the case of fully randomized experiments. While the degree of overlap between the two treatment groups can be typically examined in the data, SUTVA and unconfoundedness are untestable assumptions, and their plausibility must be justified based on domain knowledge.
	
	In this work, we focus on non-parametric regression-based approaches to CATE estimation. \cite{imbens_2004} offers a comprehensive overview on different methodologies (regression-based, matching-based, etc.) to derive different causal estimands of interest, such as sample and population ATE and CATE, Average Treatment effect on the Treated (ATT), Conditional Average Treatment effect on the Treated (CATT) etc. A non-parametric regression approach entails modelling the response surface as an unknown function of the covariates and treatment assignment indicator, and an error term. As typically done in the vast majority of the contributions on regression-based CATE estimation, we assume that the error term is additive and normally distributed with zero mean, such that $Y_i$ is modelled as
	\begin{equation} \label{onestage}
	Y_i = f(\bm{X}_i, Z_i) + \varepsilon_i~, \qquad \text{where} \quad \varepsilon_i \sim  \mathcal{N} (0,\sigma^2)
	\end{equation}
	and where $f(\cdot)$ is of unknown form, and learnt from the data. A broad variety of methods to retrieve a CATE estimator from equation (\ref{onestage}) have been developed in the literature (see \cite{caron_2020} for a review). Among these, we will follow in particular the one presented in \cite{hahn_2020}, that we introduce and discuss in the next section.

	\section{BART for causal inference} \label{sec:BARTCI}
	
	Bayesian Additive Regression Trees (BART) are a non-parametric regression model that estimates the conditional expectation of a response variable $Y_i$ via a ``sum-of-trees". Considering the regression framework in (\ref{onestage}), one can use BART to flexibly represent $f(\cdot)$ as:
	\begin{equation}
	f(\bm{X}, Z) = \sum^{m}_{j=1} g_j \Big( \big[ \bm{X}~~Z \big], ~ \big(T_j, M_j \big) \Big) ~ ,
	\end{equation}
	where $m$ is the total number of trees in the model; the pair $(T_j, M_j)$ defines the structure of the $j$-th tree, namely $T_j$ embeds the collection of binary split rules while $M_j = \{ \psi_1, ..., \psi_b \}$ the collection of $b$ terminal nodes in that tree; $g_j (\cdot)$ is a tree-specific function mapping the predictors $[\bm{X}~~Z]$ to the set of terminal nodes $M_j$, following the set of binary split rules expressed by $T_j$. The conditional mean function $f(\bm{x}, z) = \mathbb{E} \big[ Y_i \mid \bm{X}_i = \bm{x}_i, Z_i = z_i \big] $ fit is computed by summing up all the terminal nodes $\psi_{ij}$ assigned to the predictors $[\bm{X}~~Z]$ by the tree functions $g_j(\cdot)$, i.e.~$\sum^{m}_{j=1} g_j(\cdot)$. We refer the reader to \cite{cgm_1998} and \cite{chipman_2010} for more details about BART priors and inference.

	\subsection{Bayesian Causal Forests}
	
	As briefly mentioned in Section \ref{sec:probframe}, we will follow the representation proposed by \cite{hahn_2020}, that avoids imposing direct regularization on $f(\cdot)$ in (\ref{onestage}). \cite{hahn_2018} and \cite{hahn_2020} in fact show that regularization on $f (\cdot)$ can generate unintended bias in the final estimation of CATE, and propose a simple reparametrization of (\ref{onestage}) that utilizes a two-stage regression approach that dates back to the early contributions of \cite{heckman_1979} and \cite{robinson_1988}. The two-stage representation reads:
	\begin{align}
	Z_i & \sim \text{Bernoulli} \big( \pi(\bm{\tilde{X}}_i) \big) ~ , ~  \pi(\bm{\tilde{x}}_i) = \mathbb{P} (Z_i = 1 \mid \bm{\tilde{X}}_i = \bm{\tilde{x}}_i)  ~ , \label{stage1} \\
	Y_i & = \mu \Big( \big[ \bm{X}_i ~~ \pi (\bm{\tilde{X}}_i) \big] \Big) + \tau (\bm{W}_i) Z_i  + \varepsilon_i \label{stage2} ~ .
	\end{align}
	The first stage (\ref{stage1}) deals with propensity score estimation, for which any probabilistic classifier is suitable (e.g.~logistic regression, Probit BART, neural nets, etc.). In the simulated experiments shown in later later sections, we will specifically employ either default probit BART or a one-hidden-layer neural network. In general, it is advisable not to rely on aggressive regularization in the estimation of $\pi(\cdot)$, as this could accidentally result into one or more main confounders being over-shrunk and/or left out of the model. The second stage (\ref{stage2}) estimates the prognostic score $\mu (\cdot)$, defined as the effect of the covariates $\bm{X}_i \in \mathcal{X}$ on the outcome $Y_i$ in the absence of treatment $\mu (\bm{x}_i) = \mathbb{E} \big[ Y_i \mid \bm{X}_i = \bm{x}_i, Z_i = 0 \big]$, and CATE $\tau (\cdot)$. Note that we use slightly different notation for the covariates in $\mu(\cdot)$, $\tau(\cdot)$ and $\pi(\cdot)$. This is to highlight the fact that the set of available covariates $\bm{X}_i \in \mathcal{X}$ might consist of four different types \citep{herren_2020}: i) \textbf{confounders}, i.e.~direct and indirect common causes of $Z$ and $Y$; ii) \textbf{prognostic covariates}, i.e.~predictors of $\mu(\cdot)$ only; iii) \textbf{moderators}, i.e.~predictors of $\tau(\cdot)$ only; iv) \textbf{propensity covariates}, entering only $\pi(\cdot)$ equation. Any covariate that does not fall into one of these categories is an irrelevant/nuisance predictor.
	
	The two-stage procedure described above belongs to a class of models known as ``modularized", as opposed to joint-models, that attempt to embed uncertainty around propensity scores in a single stage, which nonetheless can lead to poor estimates due to feedback issues in the approximation of the full posterior \citep{zigler_2013, zigler_2014}. See \cite{jacob_2017} for a thorough discussion on the issue of modularized versus joint models.
	
	A Bayesian Causal Forest model \citep{hahn_2020} is based on the reparameterization of the second stage regression (\ref{stage2}). The advantage of this reparametrization from a Bayesian standpoint lies in the fact that separate priors, offering targeted regularization, can be placed on the prognostic score $\mu(\cdot)$ and on CATE $\tau (\cdot)$ directly. This approach mitigates unintended bias attributable to what the authors call \emph{Regularization Induced Confounding} (RIC). The intuition behind RIC is that CATE posterior is strongly influenced by the regularization effects of the prior on $f(\cdot)$ in (\ref{onestage}), such that posterior estimates of CATE are bound to be biased, even more so in presence of strong confounding, such as when treatment selection is suspected to be ``targeted'', i.e., when individuals are selected into treatment based on the prediction of an adverse potential outcome if left untreated. In order to alleviate confounding from targeted selection, the authors suggest to employ propensity score estimates obtained from the first stage $\hat{\pi}$ as an additional covariate in the estimation of $\mu (\cdot)$.
	
	In practice, a BCF model assigns a default BART prior to $\mu(\cdot)$, while a prior with stronger regularization is chosen for $\tau(\cdot)$, as moderating patterns are believed to be simpler. The BART prior on $\tau(\cdot)$, compared to the default specification, consists in the use of a smaller number of trees in the ensemble (50 trees instead of 200), and a different combination of hyperparameters that govern the depth of each tree. In particular, in the context of BART priors, the probability that a node at depth $d \in \{0, 1, 2, ...\}$ in a tree is non-terminal is given by $\nu (1 + \beta)^{-d}$, where $(\nu, \beta)$ are the hyperparameters to set \citep{chipman_2010}. The default specification $(\nu, \beta) = (0.95, 2)$ already has a shrinkage effect that accommodates small trees. The BCF prior on $\tau(\cdot)$ instead sets $(\nu, \beta) = (0.25, 3)$, with the purpose of assigning higher probability mass to even smaller trees. This combination of hyperparameters in the CATE prior allows to detect weak heterogeneous patterns, and provides robustness in case of homogeneous treatment effects.
	
	For the reasons illustrated above, BCF tends to outperform BART and other tree-based methods for CATE estimation, such as Causal Forests \citep{athey_2019}. As we will illustrate in the following sections, our work extends the BCF framework by introducing explicit shrinkage of irrelevant predictors, which results into higher computational efficiency, and accommodates different levels of smoothness across covariates, while, at the same time, returning interpretable measures of feature importance in the estimation of $\mu(\cdot)$ and $\tau (\cdot)$, separately.

	\section{Shrinkage Bayesian Causal Forests} \label{sec:sparsetree}
	
	BART, and consequently BCF, are known to handle sparsity quite well, thanks to the fact that splitting variables are chosen uniformly at random. However, they do not actively implement heterogeneous sparsity, nor feature shrinkage, which inevitably implies assigning equal level of heterogeneity to every covariate in the model. We will briefly illustrate in this section the concept of feature shrinkage in the context of tree ensemble models such as BART. Let us define first $ \bm{s} = (s_1, ..., s_P)$ as the vector of splitting probabilities of each predictor $j \in \{1, ..., P \}$, where each $s_j$ represents the probability for the $j$-th predictor of being chosen as a splitting variable in one of the decision nodes of a tree. The default version of BART places a uniform distribution over the splitting variables, meaning that each predictor has equal chance of being picked as a splitting variable: $ s_j = P^{-1} \hspace{0.3cm} \forall j \in \{ 1,..., P \} $. As a consequence, predictors are virtually given equal prior importance in the model. A sparsity-inducing solution in this framework implies having a vector $\bm{s}$ of ``stick-breaking'' posterior splitting probabilities where ideally the entries corresponding to irrelevant predictors are near-zero, while the ones corresponding to relevant predictors are significantly higher than $P^{-1}$. Posterior splitting probabilities in this context can be intuitively viewed as a measure of variables importance \citep{breiman_2001}. A complementary, decision-theoretic interpretation of sparsity-inducing solutions in this setup is given by the posterior probabilities that a predictor $j$ appears in a decision node at least once in the ensemble. The two interpretations above (variables importance and probability of inclusion) are interchangeable and qualitatively lead to the same conclusions. In the next section we review how a simple extension of BART proposed by \cite{linero_2018_sparse} can accommodate sparse solutions as described above, and how this modified version of BART can be put to use in the context of Bayesian Causal Forests. 
	
	\subsection{Dirichlet Additive Regression Trees}
	
	Dirichlet Additive Regression Trees \citep{linero_2018_sparse}, or DART, constitute an effective and practical way of inducing sparsity in BART. The proposed modification consists in placing an additional Dirichlet prior on the vector of splitting probabilities $\bm{s}$, which triggers a consequent posterior update in the backfitting MCMC algorithm. The Dirichlet prior on $\bm{s}$ reads
	\begin{equation}
	(s_1, ..., s_P) \sim \text{Dirichlet} \left( \frac{\alpha}{P}, ..., \frac{\alpha}{P} \right) ~ ,
	\end{equation}
	where $\alpha$ is the hyperparameter governing the a priori preference for sparsity. Lower values of $\alpha$ correspond to sparser solutions, that is, fewer predictors included in the model. The hyperparameter $\alpha$ is in turn assigned a prior distribution, in order to deal with unknown degree of sparsity. This prior is chosen to be a $\text{Beta}$ distribution, placed over a standardized version of the $\alpha$ parameter, of the following form
	\begin{equation}  \label{hyperprior}
	\frac{\alpha}{\alpha + \rho} \sim \text{Beta} (a, b) ~ ,
	\end{equation}
	where the default parameter values are $(a,b, \rho) = (0.5, 1, P)$. The combination of values $a=0.5$ and $b=1$ assigns higher probability to low values of $\alpha$, thus giving preference to sparse solutions (the combination $(a,b)=(1,1)$ would instead revert back to default BART splitting probabilities, i.e.~uniform distribution over the splitting variables). The prior is assigned to the standardized version of $\alpha$ in (\ref{hyperprior}) instead of $\alpha$ directly, as this allows to easily govern preference for sparsity through the parameter $\rho$. If one suspects that the level of sparsity is, although unknown, rather high, setting a smaller value of $\rho$ facilitates even sparser solutions.
	
	The modified version of DART's MCMC implies an extra step to update $\bm{s}$, according to the conjugate posterior
	\begin{equation} \label{diriposte}
	s_1, ..., s_P \mid (u_1, ..., u_P) \sim \text{Dirichlet} \left( \frac{\alpha}{P} + u_1, ..., \frac{\alpha}{P} + u_P \right) ~ ,
	\end{equation} 
	where the update depends on $u_j$, defined as the number of attempted splits on the $j$-th predictor in the current MCMC iteration. The phrase ``attempted splits" refers to the fact that BART MCMC algorithm generates trees through a branching process undergoing a Metropolis-Hastings step, so that a proposed tree in the process might be rejected, but the chosen splitting variables are counted anyway in $\bm{u} = (u_1, ..., u_P)$ \citep{cgm_1998, chipman_2010, linero_smooth}.
	
	The rationale behind the update in (\ref{diriposte}) follows the natural Dirichlet-Multinomial conjugacy. The more frequently a variable is chosen for a splitting rule in the trees of the ensemble in a given MCMC iteration (or equivalently the higher is $u_j$), the higher the weight given to that variable by the updated $\bm{s} \mid (u_1, ..., u_P)$ in the next MCMC iteration. Hence, the higher $s_j$, the higher the chance for the $j$-th predictor of being drawn as splitting variable from the multinomial distribution described by $\text{Multinom} \big(1, ~ \bm{s} \mid \bm{u} \big)$. This extra Gibbs step comes at negligible computational cost when compared to default BART typical running time.

	\subsection{Shrinkage BCF priors}
	
	Similarly to \cite{linero_2018_sparse}, symmetric Dirichlet priors can be straightforwardly embedded in the Bayesian Causal Forest framework to induce sparsity in the estimation of prognostic and moderating effects. Bearing in mind that, as described in the previous section, BCF prior consists in two different sets of independent BART priors, respectively placed on the prognostic score $\mu (\cdot)$ and CATE $\tau (\cdot) $, our proposed extension implies adding an additional Dirichlet prior over the splitting probabilities to these BART priors. Throughout the rest of the work we will consider the case where $\bm{W}_i = \bm{X}_i$, i.e.~where the same set of covariates is used for the estimation of $\mu (\cdot)$ and $\tau (\cdot) $ (see eq.\,(\ref{stage2}) for reference), but the ideas easily extend to scenarios where a different set of covariates is designed, based on domain knowledge, to be used for $\mu (\cdot)$ and $\tau (\cdot) $\footnote{In certain cases, the set of pre-treatment covariates might benefit from an initial screening by the researcher in the design of the study, and later undergo feature shrinkage in Shrinkage BCF, with the possibility of incorporating further a priori knowledge through the prior distributions, as described later in this section. As we will show in Section \ref{sec:toyexe1}, in fact, Shrinkage BCF not only adjusts to sparse data generating processes (DGPs) per se, but allocates splitting probabilities in a more efficient way among the covariates, compared to uniformly at random splits, increasing computational efficiency.}. The additional priors are respectively 
	\begin{equation}  \label{SparseBCFpriors} \small
	\begin{aligned}[c]
	\bm{s}_{\mu} & \sim \text{Dirichlet} \left( \frac{\alpha_{\mu}}{P+1}, ..., \frac{\alpha_{\mu}}{P+1} \right) ~ , \\[7pt]
	\bm{s}_{\tau}  & \sim \text{Dirichlet} \left( \frac{\alpha_{\tau}}{P}, ..., \frac{\alpha_{\tau}}{P} \right) ~ , \\
	\end{aligned}
	\qquad \quad 
	\begin{aligned}[c]
	\frac{\alpha_{\mu}}{\alpha_{\mu} + \rho_{\mu}} & \sim \text{Beta} (a, b) \\[7pt]
	\frac{\alpha_{\tau}}{\alpha_{\tau} + \rho_{\tau}} & \sim \text{Beta} (a, b) ~ , \\
	\end{aligned} 
	\end{equation} 
	where the $\text{Beta}$'s parameters are chosen to be $(a, b) = (.5, 1)$ as default. The hyperparameter $\rho$ is set equal to $(P+1)$ in the case of the prognostic score ($\rho_{\mu} = P+1$) since, when estimating $\mu (\bm{x}_i)$, we make use of $P$ covariates plus an estimate of the propensity score $\widehat{\pi} (\bm{x}_i)$ as an additional covariate. In the case of $\tau (\bm{x}_i)$, we set it equal to $\rho_{\tau} = \frac{P}{2}$ to give preference to even more targeted shrinkage, as the CATE is typically believed to display simple heterogeneity patterns and a higher degree of sparsity compared to the prognostic score.

	\begin{algorithm}[t]  \small
		\SetAlgoLined
		\DontPrintSemicolon 
		
		\KwIn{Data $(X, Z, Y)$}
		\KwOut{MCMC samples of $\left\{ \mu^{(b)}(\cdot), \tau^{(b)}(\cdot), ( \bm{s}_{\mu} \mid \bm{u}_{\mu} )^{(b)} , ( \bm{s}_{\tau} \mid \bm{u}_{\tau} )^{(b)}, \sigma^{(b)} \right\}^{B}_{b=1}$}
		\For{$b = 1, ..., B$}{
			
			\KwResult{Sample $\mu^{(b)} (\bm{x}), ( \bm{s}_{\mu} \mid \bm{u}_{\mu} )^{(b)}$}
			\For{$j=1,..., m_{\mu}$}{
				Sample tree structure $T_j{\mu} \sim p( T_j | R_j, \sigma ) \propto p(T_j) p(R_j | T_j, \sigma ) $ \;
				Sample terminal nodes $M_j{\mu} \sim p(M_j | T_j, R_j, \sigma)$ ~ (conjugate normal)
			}
			Sample $(\bm{s}_{\mu} \mid \bm{u}_{\mu}) \sim \mathcal{D} \big( \alpha_{\mu} / (P+1) + u_{1 \mu}, ~ ... ~ , \alpha_{\mu} / (P+1) + u_{(P+1) \mu} \big)$
			\;
			\;
			\KwResult{Sample $\tau^{(b)} (\bm{x}), ( \bm{s}_{\tau} \mid \bm{u}_{\tau} )^{(b)}$}
			\For{$j=1,..., m_{\tau}$}{
				Sample tree structure $T_j{\tau} \sim p( T_j | R_j, \sigma ) \propto p(T_j) p(R_j | T_j, \sigma ) $ \;
				Sample terminal nodes $M_j{\tau} \sim p(M_j | T_j, R_j, \sigma)$ ~ (conjugate normal)
			}
			Sample $(\bm{s}_{\tau} \mid \bm{u}_{\tau}) \sim \mathcal{D} \big( \alpha_{\tau} / P + u_{1 \tau}, ~ ... ~ , \alpha_{\tau} / P + u_{P \tau} \big)$
			\;
			\;
			\KwResult{Sample $\sigma^{(b)}$}
			Sample $\sigma \sim p \big( \sigma | \widehat{\mu} (\bm{x}_i), \widehat{\tau} (\bm{x}_i), Y \big)$ 
			
		}
		
		\caption{Bayesian Backfitting MCMC in Shrinkage BCF}
		\label{algo1}
	\end{algorithm}

	We refer to this setup as Shrinkage Bayesian Causal Forest (Shrinkage BCF). Naturally, the two Dirichlet priors trigger two separate extra steps in the Gibbs sampler, implementing draws from the conjugate posteriors:
	\begin{equation}
	\begin{split}
	\bm{s}_{\mu} & \mid \bm{u}_{\mu} \sim \text{Dirichlet} \big( \alpha_{\mu} / (P+1) + u_{1 \mu}, ~ ... ~ , \alpha_{\mu} / (P+1) + u_{(P+1) \mu} \big) \\[5pt]
	\bm{s}_{\tau} & \mid \bm{u}_{\tau} \sim \text{Dirichlet} \big( \alpha_{\tau} / P + u_{1 \tau}, ~ ... ~ , \alpha_{\tau} / P + u_{P \tau} \big) ~ .
	\end{split}
	\end{equation}
	
	Shrinkage BCF's setup allows first of all to adjust to different degrees of sparsity in $\mu (\cdot)$ and $\tau (\cdot)$, and thus to induce different levels of smoothness across the covariates. Secondly, it naturally outputs feature importance measures on both the prognostic score and CATE separately, given that separate draws of the posterior splitting probabilities are returned. The raw extra computational time, per MCMC iteration, is slightly greater, albeit negligible, compared to default BCF; however, Shrinkage BCF demonstrates higher computational efficiency thanks to the fact that it avoids splitting on irrelevant covariates. Thus, it necessitate far fewer MCMC iterations to converge, and improves performance under sparse DGPs. A sketch of pseudo-code illustrating the backfitting MCMC algorithm in Shrinkage BCF can be found in Box \ref{algo1}.
	
	The Dirichlet priors in Shrinkage BCF can be also adjusted to convey prior information about the relevant covariates and their relative impact on the outcome. This can be achieved by introducing a set of scalar prior weights $\bm{k} = \{ k_1, ..., k_P \} \in \mathbb{R}^{P}_{+}$, such that
	\begin{equation} \label{eq:infoprior}
	\begin{split}  
	\bm{s}_{\mu} & \sim \text{Dirichlet} \left( k_{1 \mu} \frac{\alpha_{\mu}}{P+1}, \dots , k_{(P+1) \mu} \frac{\alpha_{\mu}}{P+1} \right) ~ , \\[5pt]
	\bm{s}_{\tau} & \sim \text{Dirichlet} \left( k_{1 \tau} \frac{\alpha_{\tau}}{P}, \dots , k_{P \tau}  \frac{\alpha_{\mu}}{P} \right) ~ .
	\end{split}
	\end{equation}
	The weights can take on different values for each covariate and can be set separately for prognostic score and CATE. If the $j$-th covariate is believed to be significant in predicting $\mu (\cdot)$, then its corresponding prior weight $k_{j \mu}$ can be set higher than the others, in order to generate draws from a Dirichlet distribution that allocate higher splitting probability to that covariate. In the simulated experiment of Section \ref{sec:simulconf} we will introduce a version of Shrinkage BCF with informative priors assigning higher a priori weight to the propensity score in $\mu \big( \bm{x}_i, \pi(\bm{x}_i) \big)$, to investigate whether this helps tackling strong confounding.

	\subsection{Targeted sparsity and covariate heterogeneity}	\label{sec:toyexe1}
	
	As a result of a fully Bayesian approach to feature shrinkage, Shrinkage BCF returns non-uniform posterior splitting probabilities that assign higher weight to more predictive covariates. This automatically translates into more splits along covariates with higher predictive power, compared to default BCF. To investigate whether this more strategic allocation of splitting probabilities in Shrinkage BCF leads to better performance, we test it against a default version of BCF including all the covariates and a version of BCF that already employs the subset of relevant covariates only. Think of the latter as a sort of ``oracle" BCF that knows a priori the subset of relevant covariates, but may not assign different weights to them in terms of relative importance in the estimation of $\mu (\cdot)$ and $\tau (\cdot)$ respectively. To this end, we run a simple simulated example with $P=10$ correlated covariates, of which only $5$ are relevant, meaning that they exert some effect on the prognostic score or on CATE. We compare default BCF, ``oracle" BCF using only the $5$ relevant covariates and Shrinkage BCF using all the covariates ($5$ relevant and $5$ nuisance). We generate the $P=10$ covariates from a multivariate Gaussian $(X_1, ..., X_{10}) \sim \mathcal{N} (\bm{0}, \Sigma)$, where the entries of the covariance matrix are such that $\Sigma_{jk} = 0.6^{|j - k|} + 0.1\mathbb{I}(j \neq k)$, indicating positive correlation between predictors. Sample size is set equal to $N = 1000$. We then generate treatment assignment as $Z_i \sim \text{Bern} \big( \pi (\bm{x}_i) \big)$, where the propensity score is 
	\begin{equation}
	\pi (\bm{x}_i) = \mathbb{P} ( Z_i = 1 \mid \bm{X}_i = \bm{x}_i ) = \Phi \big( -0.4 + 0.3 X_{i,1} + 0.2X_{i,2} \big) ~ ,
	\end{equation}
	and $\Phi (\cdot)$ is the cumulative distribution function of a standard normal distribution. The prognostic score, CATE and response $Y_i$ are respectively generated as
	\begin{align}
	\mu (\bm{X}_i) & = 3 + X_{i, 1} + 0.8 \sin (X_{i, 2}) + 0.7X_{i, 3} X_{i, 4} - X_{i, 5} ~ ,  \nonumber \\
	\tau (\bm{X}_i) & = 2 + 0.8 X_{i, 1} - 0.3 X_{i, 12}^2 ~ , \\
	Y_i ~ & = \mu (\bm{x}_i) + \tau (\bm{x}_i) Z_i + \varepsilon_i ~ , \quad \text{where} \quad \varepsilon_i \sim \mathcal{N} (0, 1 ) ~ . \nonumber
	\end{align}
	In this experiment only the first five predictors are relevant. Table \ref{tab:1} shows performances of the default BCF, ``oracle" BCF run on just the $5$ relevant predictors (oracle BCF-$5$) and Shrinkage BCF (SH-BCF), averaged over $H=500$ Monte Carlo simulations. Performance of the methods is measured through: $\text{bias}$, defined as $\mathbb{E} \big[ (\hat{\tau}_i - \tau_i ) \mid \bm{X}_i = \bm{x}_i \big]$; the quadratic loss function
	\begin{equation} \label{PEHE}
	\mathbb{E} \big[ (\hat{\tau}_i - \tau_i )^2 \mid \bm{X}_i = \bm{x}_i \big] ~ ,
	\end{equation}
	where $\hat{\tau}_i$ is the model-specific CATE estimate, while $\tau_i$ is the ground-truth CATE; and finally 95\% frequentist coverage, defined as $\mathbb{P}\big( \hat{\tau}(\bm{x_i})_{low} \leq \tau(\bm{x_i}) \leq \hat{\tau}(\bm{x_i})_{upp} \big)$, where $ \hat{\tau}(\bm{x_i})_{\{low, high\}}$ are the upper and lower bounds of 95\% credible interval around $\hat{\tau}(\bm{x_i})$, returned by the MCMC. The loss function in (\ref{PEHE}) is also known as the \emph{Precision in Estimating Heterogeneous Treatment Effects} (PEHE) from  \cite{hill_2011}. Bias, PEHE and coverage estimates are estimated by computing, for each of the $H=500$ Monte Carlo simulations, their sample equivalents
	\begin{align*} \small
	\mbox{B}\widehat{\mbox{ia}}\mbox{s}_{\tau} & = \frac{1}{N} \sum^{N}_{i=1}  \Big( \hat{\tau} (\bm{x}_i) - \tau (\bm{x}_i) \Big)  \\
	\mbox{P}\widehat{\mbox{EH}}\mbox{E}_{\tau} & = \frac{1}{N} \sum^{N}_{i=1}  \Big( \hat{\tau} (\bm{x}_i) - \tau (\bm{x}_i) \Big)^2 \\
	\mbox{Co}\widehat{\mbox{ver}}\mbox{age}_{\tau} & = \frac{1}{N} \sum^{N}_{i=1} \mathbb{I} \Big( \hat{\tau}(\bm{x_i})_{low} \leq \tau(\bm{x_i}) \leq \hat{\tau}(\bm{x_i})_{upp} \Big) ~ ,
	\end{align*}
	and then averaging these over all the simulations. More precisely, Table \ref{tab:1} reports $\text{bias}$, $\sqrt{\mbox{PEHE}}$ and coverage estimates together with 95\% Monte Carlo confidence intervals.

    \begin{table}[t]
		\caption{\small Sample average $\text{bias}$, $\sqrt{\text{PEHE}}$ and 95\% coverage for default BCF, ``oracle" BCF which uses only the 5 relevant predictors (Oracle BCF-5) and Shrinkage BCF (SH-BCF). Bold text represents better performance.}
		\label{tab:1}
		\small
		\centering
		\begin{tabular}{l|ccc}
			\toprule
			\multicolumn{1}{c}{\textbf{Model}}  &  $\text{Bias}$  & $\sqrt{\text{PEHE}}$ & 95\% Coverage \\
			\midrule
			BCF  &  0.037 $\pm$ 0.008  &  0.447 $\pm$ 0.006   &   \textbf{0.92 $\pm$ 0.01}   \\
			Oracle BCF-5  &  0.034 $\pm$ 0.008  &  0.440 $\pm$ 0.006  &  0.91 $\pm$ 0.01 \\
			SH-BCF  & \textbf{0.031 $\pm$ 0.007}  &  \textbf{0.380 $\pm$ 0.006} &  0.88 $\pm$ 0.01  \\
			\bottomrule
		\end{tabular}
	\end{table}

	Shrinkage BCF shows better performance than default BCF as well as the ``oracle" BCF version in terms of bias and $\sqrt{\text{PEHE}}$, while reports just marginally lower coverage, indicating that the method allocates ``stick-breaking'' splitting probabilities in an efficient way and necessitates fewer MCMC iterations for convergence. The intuition as to why Shrinkage BCF performs better than  ``oracle" BCF, is that its priors allow not only to split more along relevant covariates instead of irrelevant ones (which explains the advantage over BCF), but also to split more frequently along covariates that are more predictive of the outcome, resulting in higher computational efficiency. To illustrate this concept, suppose we have the following trivial linear DGP with two covariates on the same scale, $Y = 2 X_1 + X_2$. Both covariates are relevant for predicting $Y$, but $X_1$ has a relatively higher impact in magnitude. DART, and thus Shrinkage BCF, allocate more splits along the more predictive dimension $X_1$, while BART produces a similar level of splits along both $X_1$ and $X_2$ and hence requires a larger number of MCMC iterations and provides noisier estimates.

	\subsection{Targeted regularization in confounded studies}
	The parametrization in BCF, and thus in Shrinkage BCF as well, is designed to effectively disentangle prognostic and moderating effects of the covariates and to induce different levels of sparsity when estimating these effects, in contrast to other methods for CATE estimation. The purpose of this section is to briefly illustrate with a simple example how naively introducing sparsity through a model that does not explicitly guard against RIC can have a detrimental effect on CATE estimates. To this end, we simulate, for $N=1000$ observations, $P=5$ correlated covariates as $(X_1, ..., X_{5}) \sim \mathcal{N} (\bm{0}, \Sigma)$, where the entries of the covariance matrix are $\Sigma_{jk} = 0.6^{|j - k|} + 0.1\mathbb{I}(j \neq k)$. The treatment allocation, prognostic score, CATE and response $Y_i$ are then respectively generated as follows: 
	\begin{align}
	Z_i & \sim \text{Bernoulli} \big( \pi (\bm{x}_i) \big) ~ ,  \nonumber \\
	\pi (\bm{x}_i) & = \Phi \big( -0.5 + 0.4 X_{i,1} \big) ~ , \nonumber \\
	\mu (\bm{X}_i) & = 3 + X_{i, 1} ~ , \nonumber \\
	\tau (\bm{X}_i) & = 0.5 + 0.5X_{i, 2}^2 ~ , \nonumber \\
	Y_i ~ & = \mu (\bm{x}_i) + \tau (\bm{x}_i) Z_i + \varepsilon_i ~ , \quad \text{where} \quad \varepsilon_i \sim \mathcal{N} (0, 1 ) ~ . \nonumber
	\end{align}
	
	\begin{table}[t]
		\caption{\small Posterior splitting probabilities from S-Learner DART, T-Learner DART and Shrinkage BCF over the 5 available covariates. Values in bold denote which covariates receive significant chunks of splitting probability in fitting the corresponding functions, that characterize each model.}
		\label{DARTs}
		\small
		\centering
		\begin{tabular}{lccccccc}
			\toprule
			\multicolumn{2}{c}{\multirow{2}{*}{\textbf{Method}}}   &  \multicolumn{6}{c}{Variable}  \\
			
			\cmidrule{3-8}
			
			\multicolumn{2}{c}{}    &  $X_1$ & $X_2$ & $X_3$ & $X_4$ & $X_5$ & $Z$  \\
			\midrule
			
			S-DART  &  $f (\cdot)$  &  \textbf{0.12} & \textbf{0.43} & 0.00 & 0.00 & 0.00 & \textbf{0.45}  \\
			\midrule
			\multirow{2}{1.5cm}{T-DART}  &  $f_0 (\cdot)$  &   \textbf{0.29} & \textbf{0.70} & 0.01 & 0.00 & 0.00 & -  \\
			&  $f_1 (\cdot)$  &   \textbf{0.09} & \textbf{0.90} & 0.00 & 0.01 & 0.00 & - \\
			\midrule
			
			\multirow{2}{1.5cm}{SH-BCF}  &  $\mu (\cdot)$  &   \textbf{0.98} & 0.01 & 0.00 & 0.00 & 0.01 & -  \\
			&  $\tau (\cdot)$  &   0.00 & \textbf{0.96} & 0.00 & 0.03 & 0.01 & - \\
			
			\bottomrule
		\end{tabular}
	\end{table}	
	
	\noindent Notice that in this simple setup the prognostic effect is determined by the first covariate $X_{i,1}$, while the moderating effect by the second covariate $X_{i,2}$. We run CATE estimation via three different methods that make use of DART priors. The first is a ``Single-Learner" (S-Learner) that employs DART (S-DART) to fit a single surface $f(\cdot)$ and computes CATE estimates as $\hat{\tau} (\bm{x}_i) = \hat{f} (\bm{x}_i, Z_i = 1) - \hat{f} (\bm{x}_i, Z_i = 0) $. The second is a ``Two-Learner" (T-Learner) that employs DART (T-DART) to fit two separate surfaces, $f_1 (\cdot)$ and $f_0 (\cdot)$, for the two treatment groups and derives CATE estimates as $\hat{\tau} (\bm{x}_i) = \hat{f}_1 (\bm{x}_i) - \hat{f}_0 (\bm{x}_i) $. The last method is our Shrinkage BCF (SH-BCF). Each of these methods is able to account for sparsity when estimating CATE. However, the interpretation of covariate importance is very different across them, due to the way the CATE estimator is derived. In particular, as indicated by the posterior splitting probabilities of each method in Table \ref{DARTs}, S-DART fits a single surface $f (\cdot)$, where $Z$ is treated as an extra covariate, so it ends up assigning most of the splitting probability to $Z$ and then in turn to other relevant covariates. T-DART performs ``group-specific" feature shrinkage, in that it fits separate surfaces for each of the treatment groups. Although both S-DART and T-DART turn out to select the relevant covariates for the final estimation of CATE, they are unable, by construction, to distinguish between prognostic and moderating ones. Shrinkage BCF instead, thanks to its parametrization, is capable of doing so, disentangling the two effects.
	
	In Section \ref{sec:simul}, we will show that Shrinkage BCF outperforms default BCF and other state-of-the-art methods in estimating CATE under two more challenging simulated exercises. Furthermore, in the supplementary material we present results from few additional simulated experiments.

	\section{Simulated experiments}  \label{sec:simul}
	
	In this section, we report results from two simulated studies carried out to demonstrate the performance of Shrinkage BCF and its informative prior version under sparse DGPs. The first simulated study is intended to evaluate Shrinkage BCF performance compared to other popular state-of-the-art methods for CATE estimation, and to show how it scales up with an increasing number of nuisance covariates. In addition, we will also illustrate how the method returns interpretable feature importance measures, as posterior splitting probabilities on $\mu (\cdot)$ and $\tau (\cdot)$. The second simulated setup instead mimics a strongly confounded study, and is designed to show how versions of Shrinkage BCF deal with targeted selection scenarios. In the supplementary material, we present further results from four additional simulated exercises, designed to: i) study what happens with perfectly known propensity scores in confounded settings; ii) investigate computational advantage of DART priors; iii) test Shrinkage BCF's reliability under increasingly larger $P$; iv) consider different types of sparse DGPs. The \texttt{R} code implementing Shrinkage BCF is available at: \url{https://github.com/albicaron/SparseBCF}.

	\subsection{Comparison to other methods} \label{sec:simcompare}
	
	The first setup consists of two parallel simulated studies, where only the total number of predictors ($P=25$ and $P=50$) is changed. The purpose underlying this setup is to illustrate how Shrinkage BCF relative performance scales up when nuisance predictors are added and the level of sparsity increases. 
	
	For both simulated exercises, sample size is set equal to $N=1000$. In order to introduce correlation between the covariates, they are generated as correlated uniforms from a Gaussian Copula $C^{\text{Gauss}}_{\Theta} (u) = \Phi_{\Theta} \big( \Phi^{-1}(u_1), \dots , \Phi^{-1}(u_P) \big) $, where $\Theta$ is a covariance matrix such that $\Theta_{jk} = 0.3^{|j - k|} + 0.1\mathbb{I}(j \neq k) $. A 40\% fraction of the covariates is generated as continuous, drawn from a standard normal distribution $\mathcal{N} (0, 1)$, while the remaining 60\% as binary, drawn from a binomial $\text{Bin} (N, 0.3)$. Propensity score is generated as:
	\begin{equation}
	\pi (\bm{x}_i) = \mathbb{P} ( Z_i = 1 \mid \bm{X}_i = \bm{x}_i ) = \Phi \left( -0.5 + 0.2 X_{i,1} + 0.1 X_{i,2} + 0.4 X_{i,21} + \frac{\eta_i}{10} \right) ~ ,
	\end{equation}
	where $\Phi (\cdot)$ is the cumulative distribution function of a standard normal, and $\eta_i$ is a noise component drawn from a uniform $\mathcal{U} (0, 1)$. The binary treatment indicator is drawn as $Z_i \sim \text{Bernoulli} \big( \pi (\bm{x}_i) \big)$. Prognostic score and CATE functions are simulated as follows:
	\begin{align} \label{muCATE}
	\begin{split}
	\mu (\bm{x}_i) = & ~ 3 + 1.5 \sin (\pi X_{i,1}) + 0.5 (X_{i,2} - 0.5)^2 + 1.5 (2 - | X_{i,3} | ) ~ + \\
	& + 1.5 X_{i,4} (X_{i,21} + 1)  
	\\[5pt]
	\tau (\bm{x}_i) = & ~ 0.1 + |X_{i,1} - 1|  (X_{i,21} + 2) ~ .
	\end{split}
	\end{align}
	Notice that only 5 predictors among $P \in \{25, 50\}$, namely $\{X_1, X_2, X_3, X_4, X_{21}\}$, are relevant to the estimation of the prognostic score and CATE. Eventually, the response variable $Y_i$ is generated as usual:
	\begin{equation}
	Y_i = \mu (\bm{x}_i) + \tau (\bm{x}_i) Z_i + \varepsilon_i ~ , \quad \text{where} \quad \varepsilon_i \sim \mathcal{N} (0, \sigma^2 ) ~ .
	\end{equation}
	The error term standard deviation is set equal to $\sigma = \frac{\hat{\sigma}_{\mu}}{2}$, where $\hat{\sigma}_{\mu}$ is the sample standard deviation of the simulated prognostic score $\mu (\bm{x}_i)$ in (\ref{muCATE}).

\begin{table}[t]
		\caption{List of models tested on the simulated experiment in Section \ref{sec:simcompare}.}
		\label{testedmodels}
		\small
		\centering
		\begin{tabular}{lcl}
			\toprule
			\textbf{Family}     &  \textbf{Label}     &  \textbf{Description}  \\
			\midrule
			
			\multirow{3}{3cm}{Linear Models}  &  \multirow{1}{2.7cm}{\centering S-OLS}  &   Linear regression as S-Learner  \\
			&  T-OLS  &   Linear regression as T-Learner \\
			&  R-LASSO  &   LASSO regression as R-Learner \\
			\midrule

			Naive Non-Parametrics	&  $k$NN  &   $k$-Nearest Neighbors as T-Learner \\
			
			\midrule

			\multirow{7}{3.2cm}{Tree-Based Methods}  &  \multirow{1}{2cm}{\centering S-BART}  &   BART as S-Learner \\
			&  T-BART  &   BART as T-Learner \\
			&  CF  &   Causal Forest \\
			&  S-DART  &   DART as S-Learner \\
			&  T-DART  &   DART as T-Learner \\
			&  BCF  &   Bayesian Causal Forest \\
			&  SH-BCF  &   Shrinkage Bayesian Causal Forest \\
			
			\midrule
			
			\multirow{2}{3.2cm}{Gaussian Processes}  &  \multirow{1}{2cm}{\centering CMGP}  &   Causal Multi-task Gaussian Process \\
			&  NSGP  &   Non-Stationary Gaussian Process \\

			\bottomrule
		\end{tabular}
	\end{table}

	Performance of each method is evaluated through $\sqrt{\mbox{PEHE}}$ estimates, averaged over $H=1000$ replications, reported together with 95\% Monte Carlo confidence intervals. Data are randomly split in 70\% train set, used to train the models, and 30\% test set to evaluate the model on unseen data; $\sqrt{\mbox{PEHE}}$ estimates are reported both for train and test data.

	The models evaluated on the simulated data are summarized in Table \ref{testedmodels}. We make use of the Meta-Learners terminology described in \cite{kunzel_2017} and \cite{caron_2020}. The first set of models includes a S-Learner and a T-Learner least squares regressions (S-OLS and T-OLS), and a R-Learner \citep{wager_2019} LASSO regression (R-LASSO). The second set consists just in a naive $k$-nearest neighbors ($k$NN) as a T-Learner. The third set includes the following popular tree ensembles methods: Causal Forest (CF) \citep{athey_2019}; a S-Learner and a T-Learner versions of BART (S-BART and T-BART) and DART (S-DART and T-DART); Bayesian Causal Forest (BCF) \citep{hahn_2020}; and finally our method, Shrinkage Bayesian Causal Forest (SH-BCF). The last set includes two causal multitask versions of Gaussian Processes, with stationary (CMGP) and non-stationary (NSGP) kernels respectively, both implementing sparsity-inducing Automatic Relevance Determination over the covariates \citep{vanderschaar_2017, vanderschaar_2018}.

	\begin{table}[t]
		\caption{\small Train and test set $\sqrt{\mbox{PEHE}}$ estimates, together with 95\% confidence interval, in the case of $P=25$ covariates and $P=50$ covariates scenarios.}
		\label{simuresults}
		\small
		\centering
		\begin{tabular}{ccc c cc}
			\toprule
			&   \multicolumn{2}{c}{$\bm{P=25}$} & & \multicolumn{2}{c}{$\bm{P=50}$}  \\
			
			\cmidrule(r){2-3}  \cmidrule(r){5-6}
			
			& Train  &  Test &   & Train  &  Test  \\
			
			\midrule
			
			S-OLS  		&	1.91 $\pm$ 0.00		&  1.91  $\pm$ 0.01  &   &  1.91 $\pm$ 0.00  &	1.91 $\pm$ 0.01  \\
			T-OLS  		&	1.41 $\pm$ 0.01	&  1.47 $\pm$ 0.01      & 	&  1.68 $\pm$ 0.01   &	1.78 $\pm$ 0.01  \\
			R-LASSO  		&	1.17 $\pm$ 0.01 	&   1.19 $\pm$ 0.01      &	 &  1.20 $\pm$ 0.01  &	1.22 $\pm$ 0.01	\\
			
			\midrule

			$k$NN 		&	1.62 $\pm$ 0.01 	&   1.66 $\pm$ 0.01      &   &  1.72 $\pm$ 0.01  &	1.76 $\pm$ 0.01	\\
			
			\midrule
			
			S-BART  		&	0.77 $\pm$ 0.01 	&  0.79 $\pm$ 0.01      &	&  0.85 $\pm$ 0.01  &	0.86 $\pm$ 0.01	\\
			T-BART  		&	1.11 $\pm$ 0.01 	&  	1.11 $\pm$  0.01      &	 &  1.28 $\pm$ 0.01  &	1.29 $\pm$ 0.01	\\
			CF  		&	1.05 $\pm$ 0.01	&  1.05 $\pm$ 0.01      &    &  1.23 $\pm$ 0.01  &	  1.23 $\pm$ 0.01	\\
			S-DART  		&	0.59 $\pm$ 0.01 	&  0.60 $\pm$ 0.01      &	&  0.59 $\pm$ 0.01  &	0.60 $\pm$ 0.01	\\
			T-DART  		&	0.88 $\pm$ 0.01 	&  	0.89 $\pm$  0.01      &	 &  0.90 $\pm$ 0.01  &	0.90 $\pm$ 0.01	\\
			BCF 		&   0.79 $\pm$ 0.01		&     0.82 $\pm$ 0.01      &	&  0.86 $\pm$ 0.01  &	0.88 $\pm$ 0.01	\\
			\textbf{SH-BCF}  		&	\textbf{0.54 $\pm$ 0.01} 	&  	\textbf{0.56 $\pm$ 0.01}   &   &  \textbf{0.55 $\pm$ 0.01}  &  \textbf{0.55 $\pm$ 0.01}	\\
			
			\midrule
			
			CMGP  		&   0.59 $\pm$ 0.01 	&	0.61 $\pm$ 0.01      &	  &  0.85 $\pm$ 0.03 &	0.77 $\pm$ 0.02 \\
			NSGP		&  	0.60 $\pm$ 0.01 	&   0.62 $\pm$ 0.01      &   &  0.74 $\pm$ 0.03  &	0.75 $\pm$ 0.03	\\

			\bottomrule
		\end{tabular}  
	\end{table} 	
	
	Performance of each method, for the two simulated scenarios with $P=25$ and $P=50$ covariates respectively, is shown in Table \ref{simuresults}. Results demonstrate the high adaptability and scalability of Shrinkage BCF, as the method displays the lowest estimated error in both simulated scenarios, and its performance is not undermined when extra nuisance covariates are added, while the other methods generally deteriorate.

	\begin{figure}[t]
		\centering
		\begin{subfigure}{.5\textwidth}
			\centering
			\includegraphics[scale=0.49]{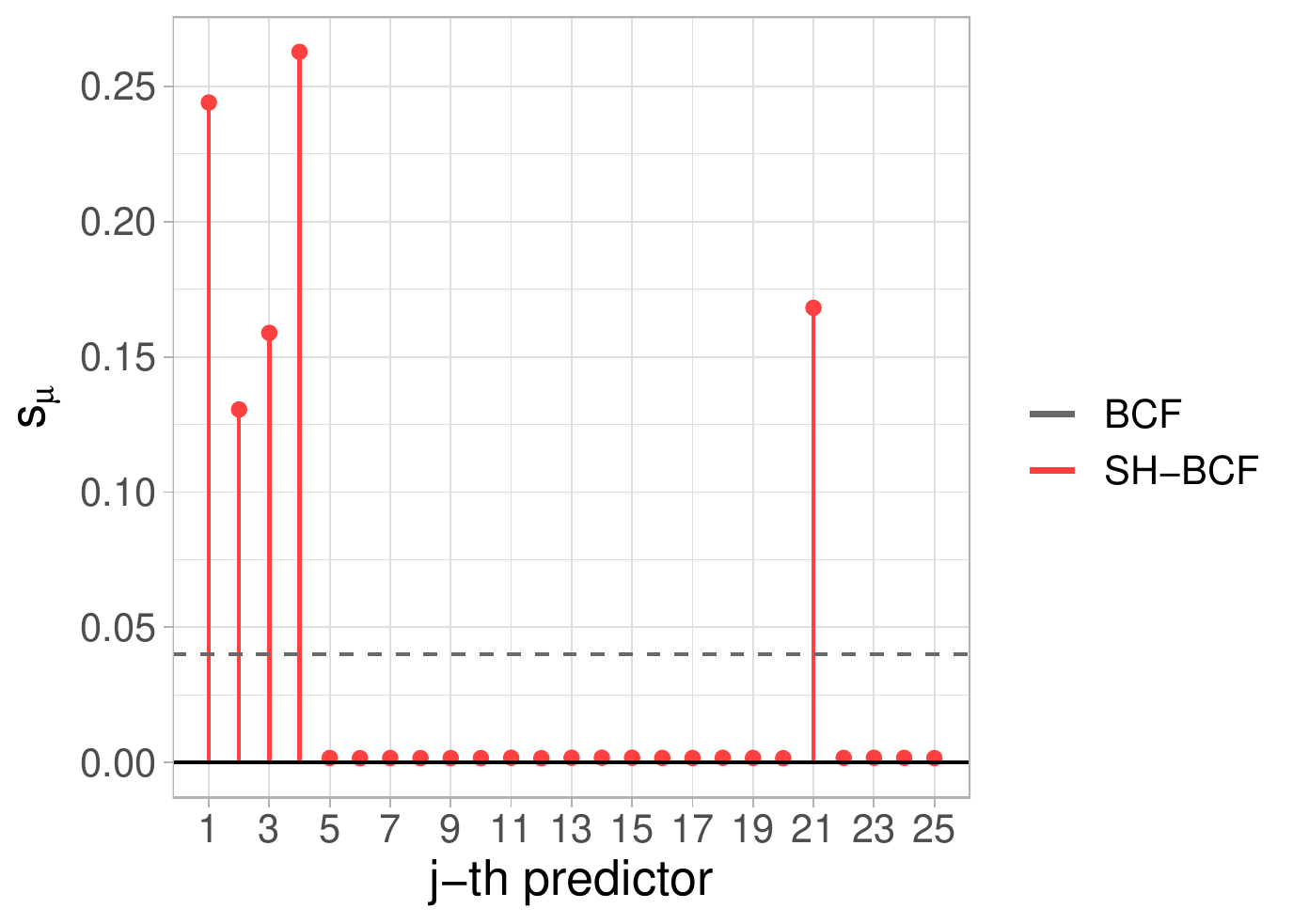}
		\end{subfigure}%
		\begin{subfigure}{.5\textwidth}
			\centering
			\includegraphics[scale=0.49]{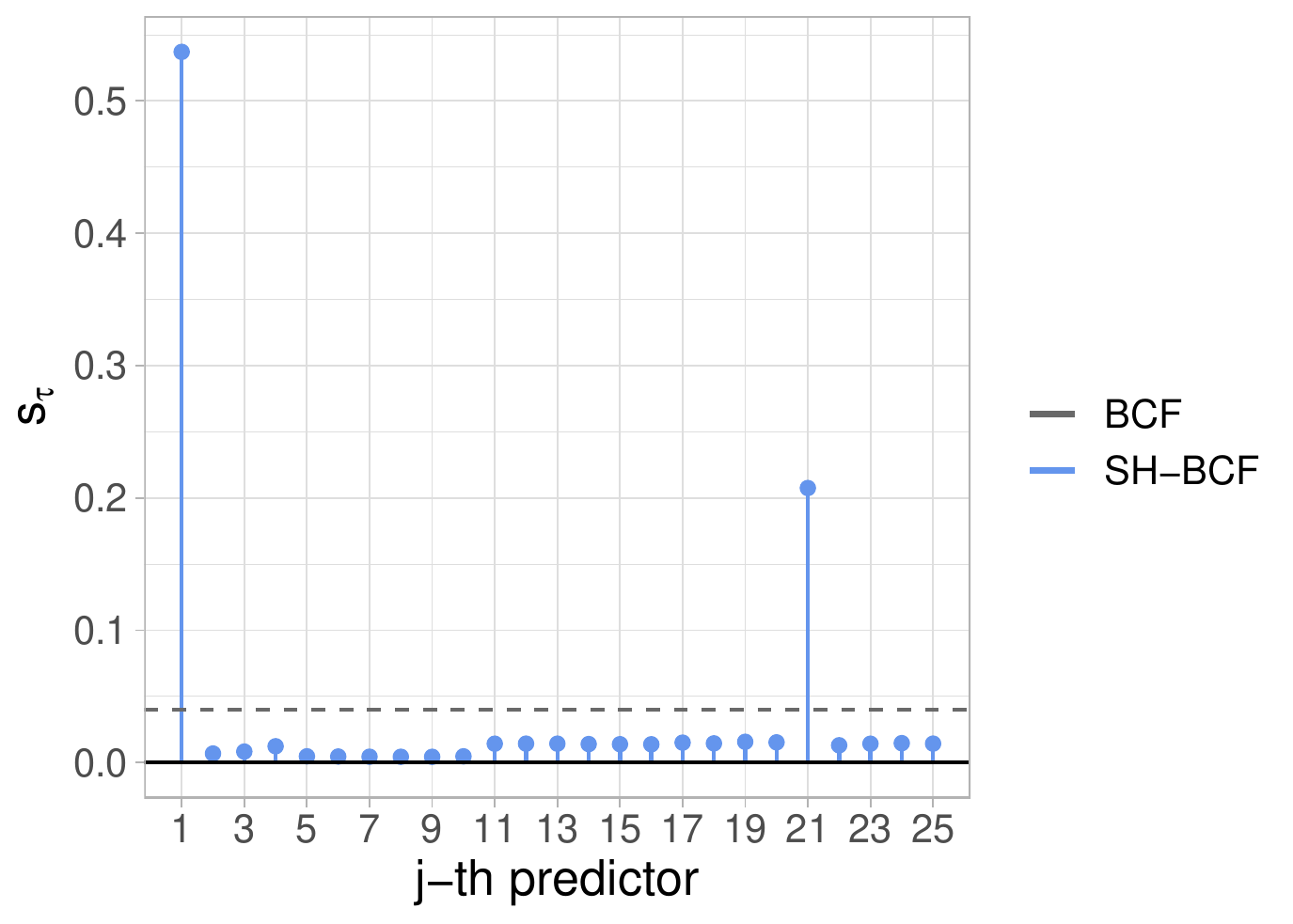}
		\end{subfigure}
		
		\begin{subfigure}{.5\textwidth}
			\centering
			\includegraphics[scale=0.49]{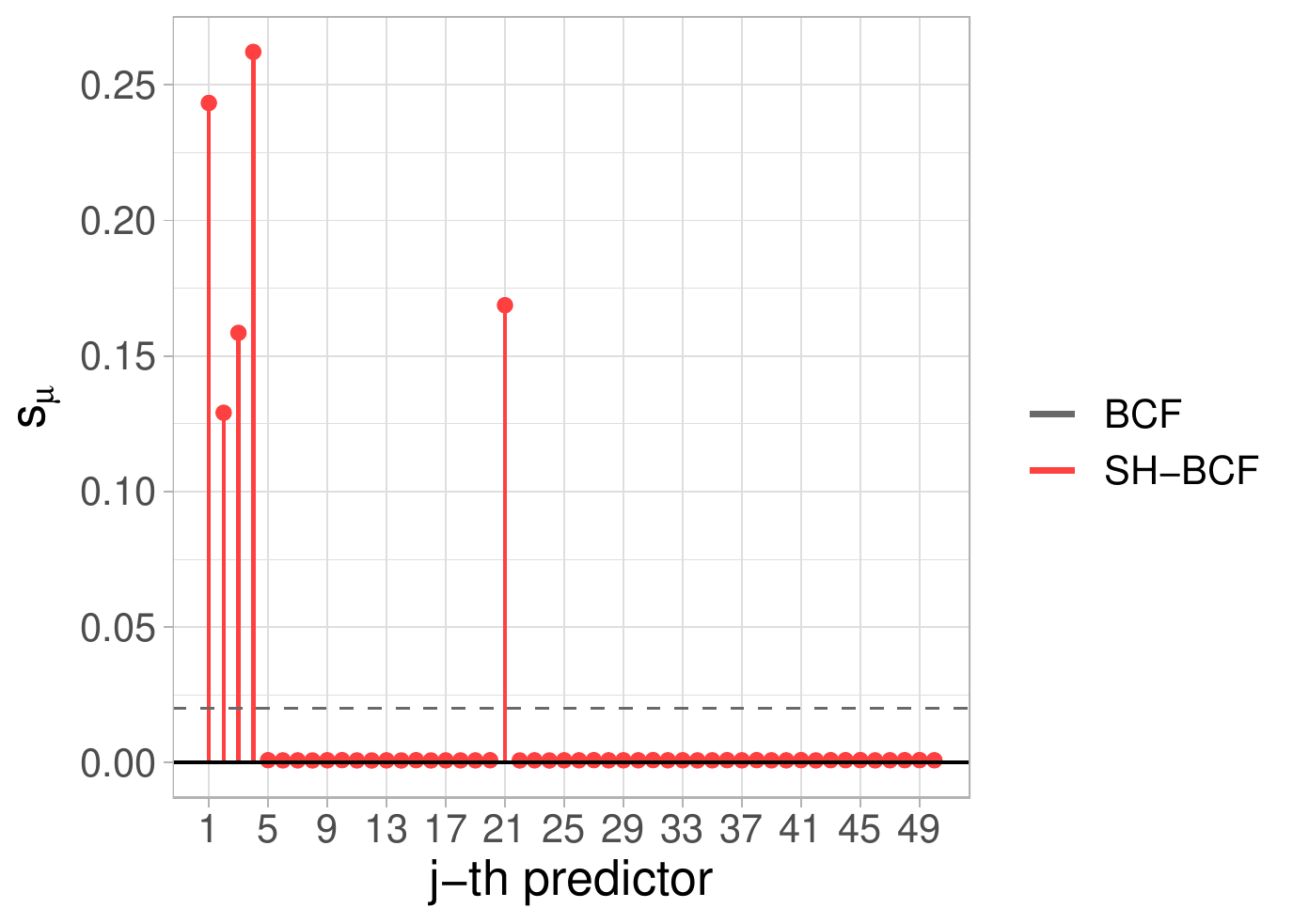}
		\end{subfigure}%
		\begin{subfigure}{.5\textwidth}
			\centering
			\includegraphics[scale=0.49]{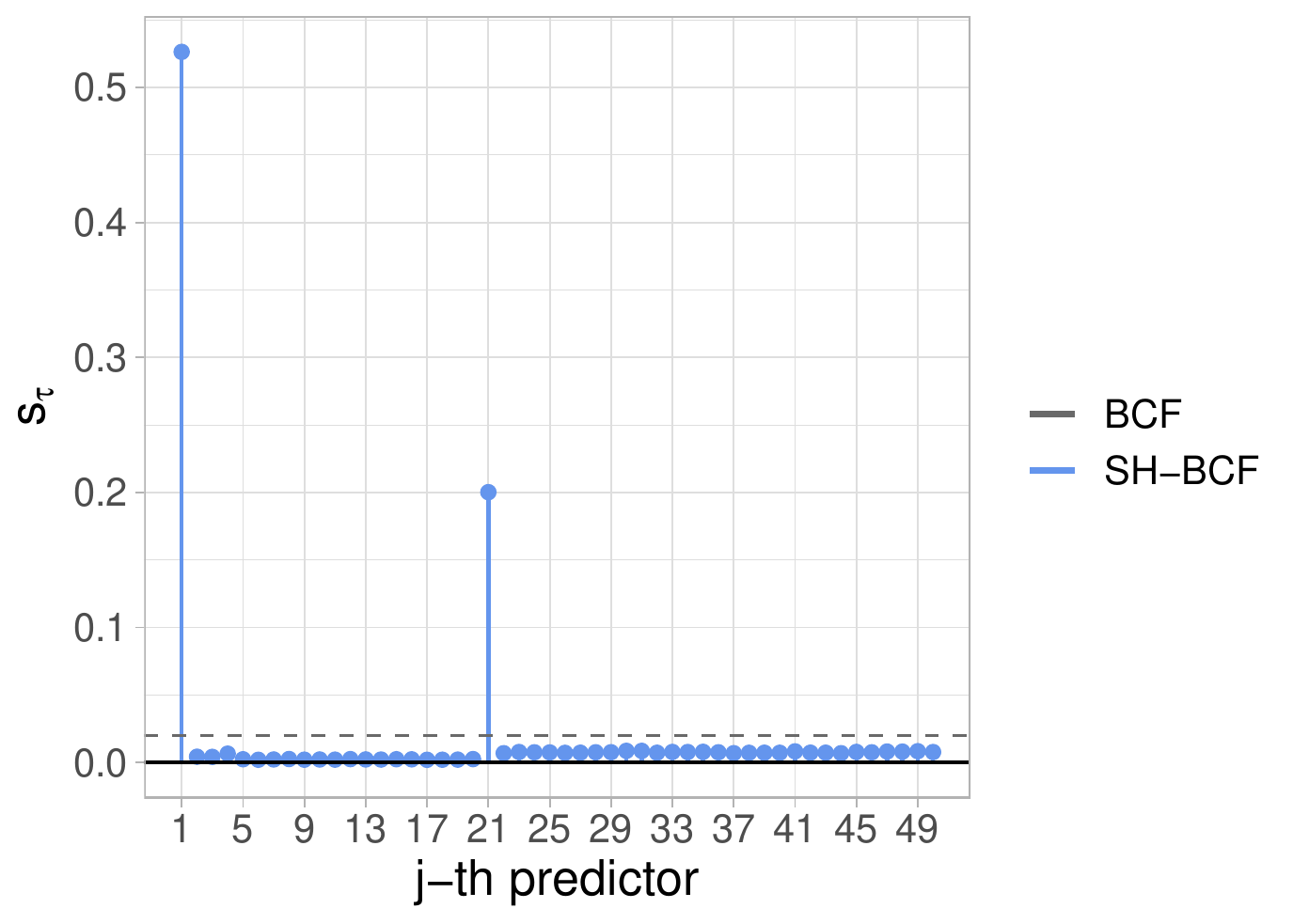}
		\end{subfigure}

		\caption{\small Shrinkage BCF posterior splitting probabilities for each single covariates, indexed on the x-axis, for $\mu (\cdot)$ (on the left) and $\tau (\cdot)$ (on the right), in the scenarios with $P = 25$ predictors (first row) and $P = 50$ predictors (second row). Spikes indicate higher probability assigned by Shrinkage BCF to the relevant predictors. The horizontal dashed lines denote default BCF uniform splitting probabilities.}
		\label{splitproba}
	\end{figure}

	Figure \ref{splitproba} shows how Shrinkage BCF correctly picks the relevant covariates behind both prognostic and moderating effects, in contrast to default BCF which assigns equal probability of being chosen as a splitting variable to each predictor. Notice also that results do not essentially vary between the $P=25$ and the $P=50$ scenarios (respectively first and second row graphs in Figure \ref{splitproba}), as Shrinkage BCF virtually selects the same relevant predictors.

	\subsection{Strongly confounded simulated study}  \label{sec:simulconf}
	
	This section presents results from a second simulated study, aimed at showing how Shrinkage BCF addresses scenarios characterized by strong confounding. In particular, the setup is designed around the concept of targeted selection, a common type of selection bias in observational studies, expressively tackled by the BCF framework, that implies a direct relationship between $\mu(\cdot)$ and $\pi(\cdot)$. We run the simulated experiment in the usual way, by firstly estimating the unknown propensity score; then we also re-run the same experiment assuming that propensity score is known (results in the supplementary material), to gain insights by netting out effects due to propensity model misspecification.
	
	We simulate $N=500$ observations from $P=15$ correlated covariates (the first 5 continuous and the remaining 10 binary), generated as correlated uniforms from the Gaussian Copula $C^{\text{Gauss}}_{\Theta} (u) = \Phi_{\Theta} \big( \Phi^{-1}(u_1), \dots , \Phi^{-1}(u_P) \big) $, where the covariance matrix is such that $\Theta_{jk} = 0.6^{|j - k|} + 0.1\mathbb{I}(j \neq k) $. The relevant quantities are simulated as follows: 
	\begin{align} 
	\mu (\bm{x}_i) = & ~ 5 \Big(2 + 0.5 \sin (\pi X_{i,1}) - 0.25 X_{i,2}^2 + 0.75 X_{i,3} X_{i,9} \Big) ~ , \nonumber \\
	\tau (\bm{x}_i) = & ~ 1 + 2 |X_{i,4}| + 1 X_{i,10} ~ , \nonumber \\
	\pi (\bm{x}_i) = & ~ 0.9 ~ \Lambda \left( 1.2 + 0.2 \mu (\bm{x}_i) \right) ~ , \label{eq:tarselstudy} \\
	Z_i \sim  & ~  \text{Bernoulli} \big( \pi (\bm{x}_i) \big) ~ ,  \nonumber \\
	Y_i =  & ~  \mu (\bm{x}_i) + \tau (\bm{x}_i) Z_i + \varepsilon_i ~ , \quad \text{where} \quad \varepsilon_i \sim \mathcal{N} (0, \sigma^2) ~ , \nonumber
	\end{align}
	where $\Lambda(\cdot)$ is the logistic cumulative distribution function. The error's standard deviation is set equal to half the sample standard deviation of the generated $\tau(\cdot)$, $\sigma^2 = \frac{\hat{\sigma}_{\tau}}{2}$. Targeted selection is introduced by generating the propensity score $\pi (\bm{x}_i)$ as a function of the prognostic score $\mu (\bm{x}_i)$ \citep{hahn_2020}. The BCF models tested on this simulated setup are: i) Default BCF; ii) agnostic prior Shrinkage BCF; iii) agnostic prior Shrinkage BCF, without propensity score estimate as an additional covariate; iv) Shrinkage BCF with informative prior on $\mu(\cdot)$ only, where prior weight given to propensity score is $k_{PS} = 50$; v) Shrinkage BCF with the same prior as iv), but $k_{PS} = 100$. We test a variety of BCF versions to examine how they tackle confounding deriving from targeted selection. In particular, with iv) and v), we investigate whether nudging more splits on the propensity score covariate induces better handling of confounding and better CATE estimates. With ii) and iii) we study whether it is sensible to have propensity score as an extra covariate, once we have accounted for sparsity, in settings such as the one described in (\ref{eq:tarselstudy}), where propensity $\pi(\cdot)$ and prognostic score $\mu(\cdot)$ are functions of the same set of covariates --- more specifically $\pi(\cdot)$ is a function of $\mu(\cdot)$.

	\begin{table}[t]
		\caption{\small $\text{Bias}$, $\sqrt{\text{PEHE}}$, 95\% Coverage and posterior splitting probability on $\hat{\pi}(x_i)$ --- $(s_{\pi} \mid u_{\pi})$ --- for: i) default BCF; ii) Shrinkage BCF; iii) Shrinkage BCF without $\hat{\pi}(x_i)$; iv) informative prior BCF with $k_{PS} = 50$; v) informative prior BCF with $k_{PS} = 100$.}
		\label{tab:tarsel}
		\small
		\centering
		\begin{tabular}{l | ccc | c}
			\toprule
			\multicolumn{1}{c}{\textbf{Model}}  &  $\text{Bias}$  & $\sqrt{\text{PEHE}}$ & 95\% Coverage & $(s_{\pi} \mid u_{\pi})$ \\
			\midrule
			i) BCF  &  -0.06 $\pm$ 0.01  &  0.49 $\pm$ 0.01  &  0.94 $\pm$ 0.00 &  9.09\% \\
			ii) SH-BCF  & -0.05 $\pm$ 0.01  &  0.38 $\pm$ 0.01 & 0.96 $\pm$ 0.00 & 0.29\% \\
			iii) SH-BCF (no PS)   &  -0.05 $\pm$ 0.01  &  0.38  $\pm$  0.01  &  0.96 $\pm$ 0.00  &  - \\
			iv) I-BCF ($k_{PS} = 50$)    &  -0.05 $\pm$ 0.01  &  0.39 $\pm$ 0.01 & 0.96 $\pm$ 0.00 &  9.76\% \\
			v) I-BCF ($k_{PS} = 100$)    &  -0.05 $\pm$ 0.01  &  0.40 $\pm$ 0.01 & 0.96 $\pm$ 0.01 &  17.48\%  \\
			
			\bottomrule
		\end{tabular}
	\end{table}
	
	We first compare the usual performance metrics (bias, $\sqrt{\text{PEHE}}$, 95\% coverage), averaged over $H=500$ replications, which are gathered in Table \ref{tab:tarsel}, together with the average posterior splitting probability assigned to propensity score ($s_{\pi} \mid u_{\pi}$) by each model, where applicable. As for the posterior splitting probability $(s_{\pi} \mid u_{\pi})$, we notice that in ii) this is nearly zero, thus not really different than not having $\pi(\cdot)$ at all, as in iii). This means that estimates of $\pi(\cdot)$ do not virtually contribute a lot to the fit. Also, in i) and iv), the probability is more or less the same, meaning that, in this example, setting $k_{PS} = 50$ implies assigning similar $(s_{\pi} \mid u_{\pi})$ as default BCF, but allowing sparsity across the other covariates. In addition to the information in Table \ref{tab:tarsel}, for a better visual inspection, we plot the posterior fit of the $\pi(\cdot)$ and $\mu(\cdot)$ relationship for each specification of BCF\footnote{We avoid plotting the fit for iii) Shrinkage BCF without $\pi(\cdot)$, since it yields very similar results to ii) Shrinkage BCF with $\pi(\cdot)$ --- In Table \ref{tab:tarsel}, ii) allocates nearly 0\% splits to $\pi(\cdot)$, as in iii).}.

	\begin{figure}[t]
		\centering
		\includegraphics[scale = 0.475]{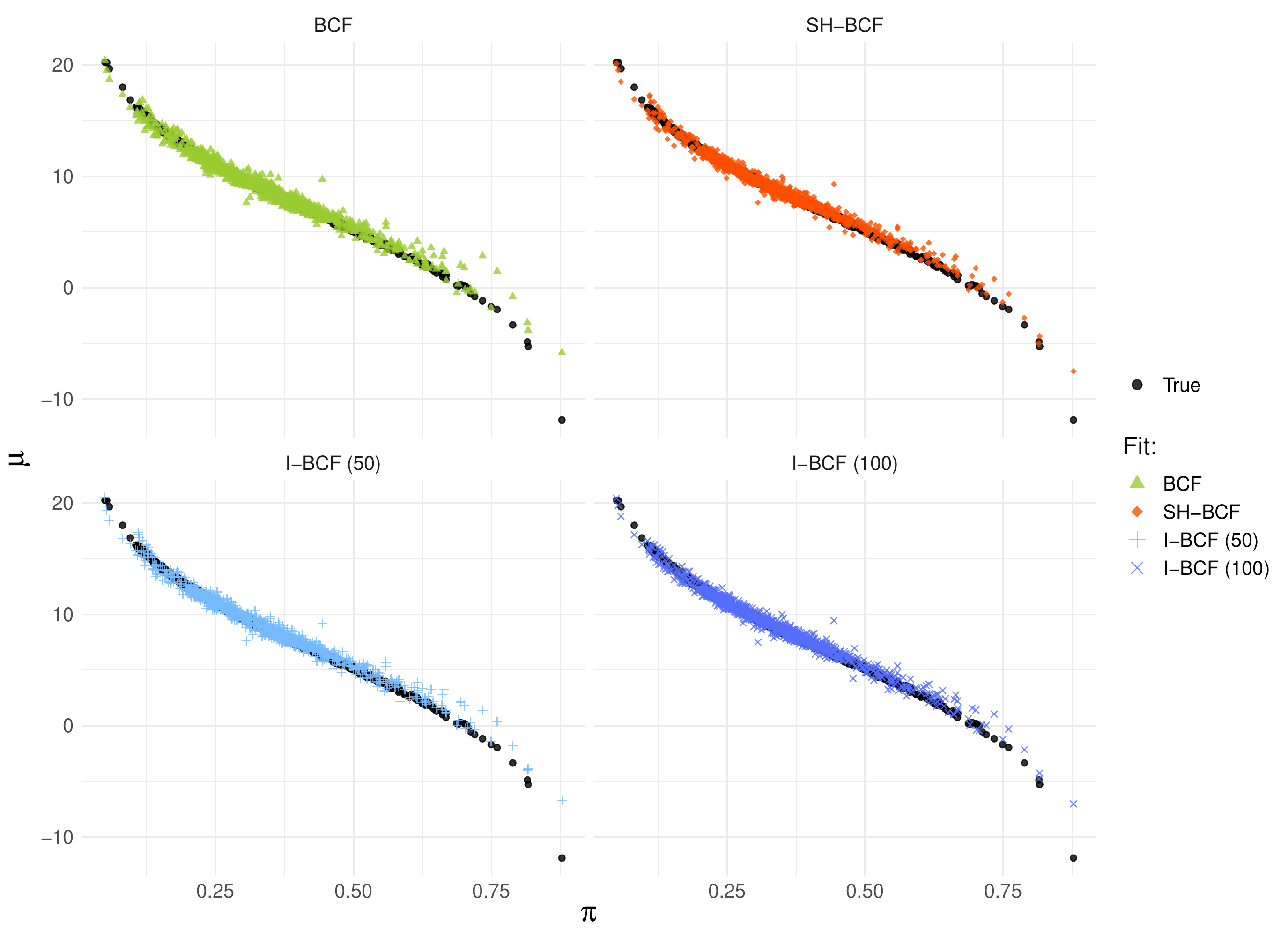}
		\caption{Posterior fit of $\pi(\cdot)$ and $\mu(\cdot)$ relationship, for default BCF, Shrinkage BCF (with $\pi(\cdot)$) and the two versions of informative prior BCF ($k_{PS} = 50$ and $k_{PS} = 100$). All the specifications effectively capture the underlying relationship.}
		\label{fig:MUvsPS}
	\end{figure}
	
	The results corroborate those of the previous sections, as all the Shrinkage BCF versions ii)-v) outperform default BCF i), thanks to their ability to adapt to sparsity (Table \ref{tab:tarsel}). In order to net out effects that are due to propensity model misspecification, we re-run the same example in (\ref{eq:tarselstudy}) for $H=250$, this time assuming that PS is known, thus plugging in the true values in $\mu(\bm{x}_i, \pi)$. Results can be found in the supplementary material.
	
	The picture emerging from this exercise is the following. Methods ii)-v) all have comparable performances in the realistic scenario where PS is to be estimated (see Table \ref{tab:tarsel}); moreover, Figure \ref{fig:MUvsPS} show that, in this case, they all effectively capture the relationship between $\pi(\cdot)$ and $\mu(\cdot)$. Hence, adjusting prior weights to nudge more splits on the estimated PS --- methods iv) and v) --- does not seem to improve performance. In the more abstract scenario where PS is assumed to be known (whose results are gathered in supplementary material), and thus the relationship between $\pi(\cdot)$ and $\mu(\cdot)$ can be directly estimated, versions i) and iii) perform poorly. The first because it does not induce sparsity, while iii) does not include $\pi(\cdot)$ as extra covariate. Versions ii), iv) and v) instead perform comparatively better as they virtually assign all the splitting probability to $\pi(\cdot)$, leaving the other covariates out of the model. This is unsurprising in a setup where $\pi(\cdot)$ is known, as its relationship with $\mu(\cdot)$ is straightforwardly captured. Even under this abstract scenario, specifications iv) and v), which assign higher weight to $\pi(\cdot)$, do not show improvements on performance, as also the agnostic prior version ii) effectively allocates the entire splitting probability to the $\pi(\cdot)$ covariate. 
	
	Results from the example where PS is perfectly known are in line with the findings of \cite{hahn_2020} and shed light on why adding $\pi(\cdot)$ as an extra covariate is always helpful in tackling targeted selection. Naturally, the success of this practice in addressing strong confounding heavily depends on the quality of the approximation of $\pi(\cdot)$, that is, the quality of the propensity model that estimates $\hat{\pi}(\cdot)$.

	\section{Case study: the effects of early intervention on cognitive abilities in low birth weight infants}  \label{sec:realworld}
	
	In this section, we illustrate the use of Shrinkage BCF by revisiting the study in \cite{brooks_1992}, which analyzes data from the Infant Health and Development Program (IHDP), found also in the more recent contribution of \cite{hill_2011}. The IHDP was a randomized controlled trial aimed at investigating the efficacy of educational and family support services, with pediatric follow-ups, in improving cognitive skills of low birth weight preterm infants, who are known to have developmental problems regarding visual-motor and receptive language skills \citep{mccormick_1985, mccormick_1990}. The study includes observations on 985 infants whose weight at birth was less than $2\,500$ grams, across 8 different sites. About one third of the infants were randomly assigned to treatment ($Z_i = 1$), which consisted in routine pediatric follow-up (medical and developmental), in addition to frequent home visits to inform parents about child's progress and communicate instructions about recommended activities for the child. Following \cite{hill_2011}, the outcome variable ($Y_i$) we use is the score in a Stanford Binet IQ test, whose values can range from a minimum of 40 to a maximum of 160, taken at the end of the intervention period (child's age equal 3). The available final sample, obtained after removing 77 observations with missing IQ test score, consists of $N = 908$ data points, while the number of pre-treatment covariates amounts to $P = 31$. A full list of the variables included in the analysis, together with a short description, can be found in the supplementary material.
	
\begin{figure}[t]
		\centering
		\includegraphics[scale = 0.482]{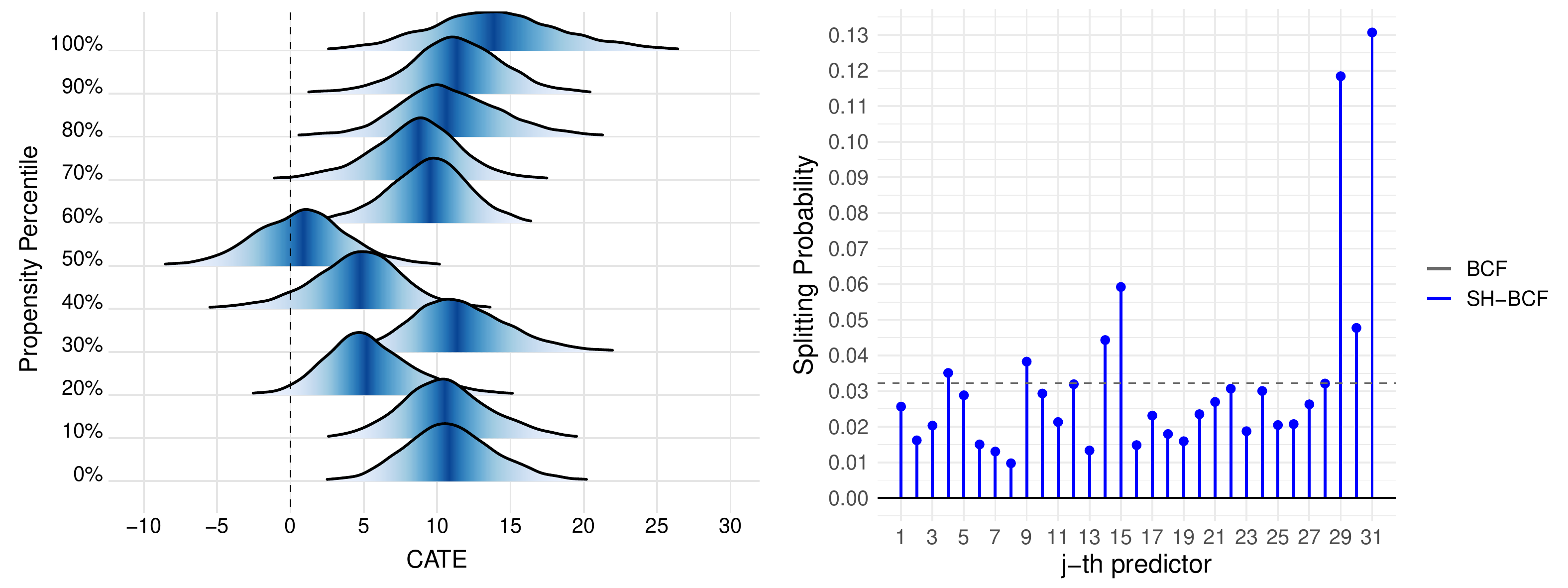}
		\caption{Left panel: Posterior distributions for the CATE estimates, obtained using Shrinkage BCF, corresponding to the approximated propensity percentiles (i.e.~for individuals in the sample whose estimated propensity corresponds or is closest to the percentiles). Fill colour is darker around the median. Right panel: Shrinkage BCF's posterior splitting probabilities on $\tau(\cdot)$, averaged over the post burn-in MCMC draws.}
		\label{fig:CATEsplit}
	\end{figure}

	Firstly, we estimate propensity score using a 1-hidden layer neural network classifier. Then we run Shrinkage BCF with default agnostic prior for $15\,000$ MCMC iterations in total, but we discard the first $10\,000$ as burn-in. As output, we obtain the full posterior distribution on CATE estimates and splitting probabilities relative to each covariate. The left-hand pane graph of Figure \ref{fig:CATEsplit} shows the estimated CATE posterior distribution for the individuals in the sample whose estimated propensity corresponds, or is closest, to the $i$-th percentile of the estimated propensity distribution, where $i$ is 0, 10, 20, ..., 100. The represented stratified CATE posterior distribution relative to these propensity values conveys information about the uncertainty around the estimates and depicts an overall positive and rather heterogeneous treatment effects. The estimated average treatment effect is equal to $\text{ATE} = 9.33$ and standard deviation of CATE estimates, averaged over the post burn-in draws, is equal to 3.25, which is another sign of underlying heterogeneity patterns in the treatment response. The analysis would thus benefit from further investigation about the heterogeneity of treatment effects, with the aim of distinguishing the impact within subgroups of individuals characterized by similar features (i.e.~covariates values). Evidence on what the relevant drivers of heterogeneity behind treatment effect are is given by the posterior splitting probabilities on $\tau(\cdot)$ (again averaged over the post burn-in draws), reported in the right-hand pane graph of Figure \ref{fig:CATEsplit}, where few covariates end up being assigned relatively higher weights compared to the others. The two covariates that primarily stand out are the binary indicator on whether the mother's ethnicity is white (29$^{th}$ predictor) and the ordinal variable indicating mother's level of education (31$^{st}$ predictor).
	
	We proceed with a sensitivity analysis of treatment effect subgroups by following the suggestion of \cite{hahn_2020}; that is, we fit a decision tree partition algorithm using the \texttt{R} package \texttt{rpart}, by regressing mean CATE estimates obtained from Shrinkage BCF $\hat{\tau}(\bm{x}_i)$ (averaged over the MCMC post burn-in draws) on the available covariates $X_i \in \mathcal{X}$. The purpose of this sensitivity analysis exercise is to identify the most homogeneous subgroups, namely the subgroups leading to an optimal partition, in terms of their estimated mean CATE, as a function of the covariates, and to examine how much the emerging partition agrees with the results on posterior splitting probabilities in Figure \ref{fig:CATEsplit}. 
	
\begin{figure}[t]
		\centering
		\includegraphics[scale = 0.53]{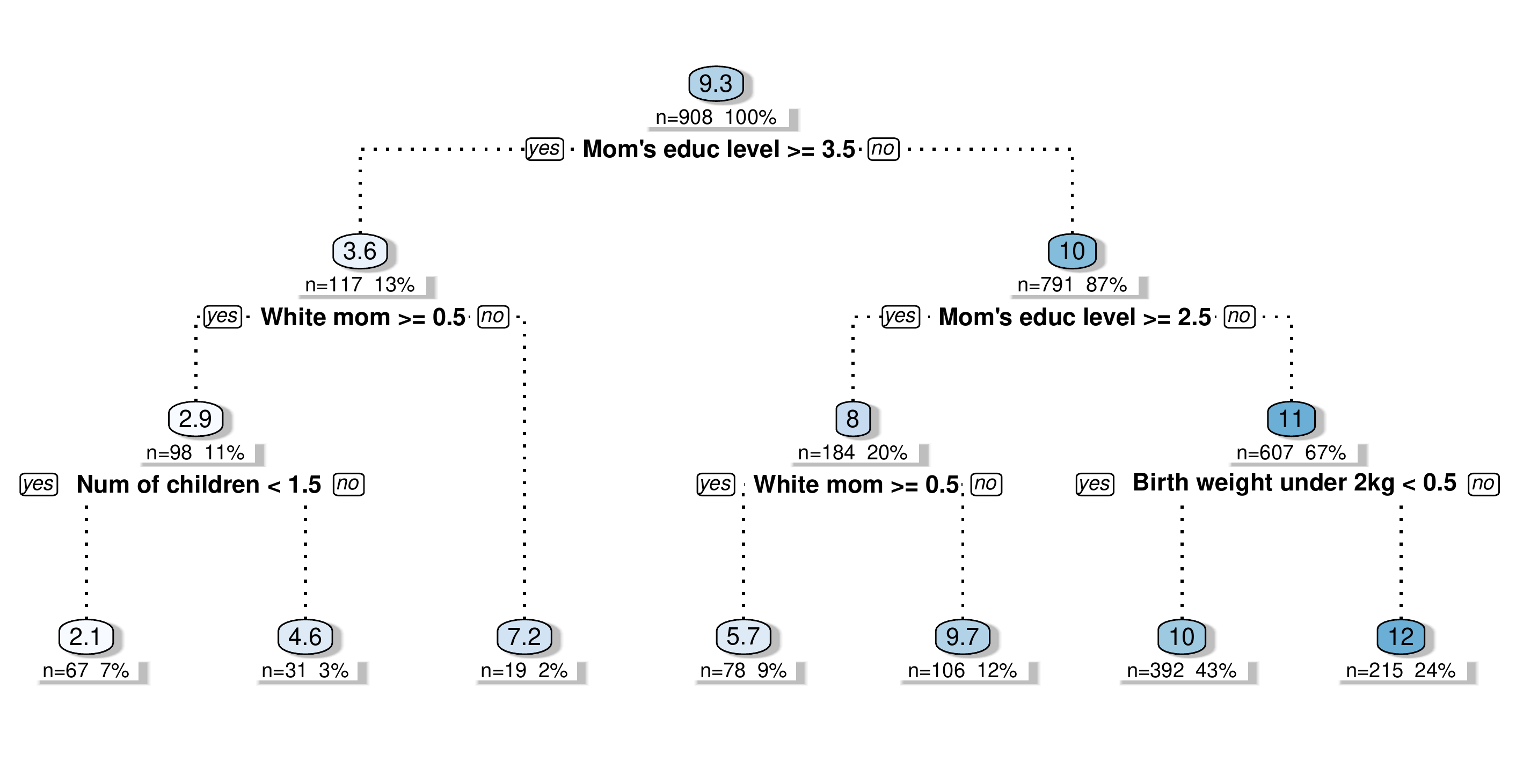}
		\caption{Decision tree identifying the most homogeneous subgroups in terms of treatment response, based on splitting rules involving the available covariates. The nodes report CATE estimates averaged within the corresponding subgroup.}
		\label{fig:dectree}
	\end{figure}

	Results are depicted in Figure \ref{fig:dectree} in the form of a decision tree, pruned at four levels. Zero splits trivially return ATE estimate (first node in Figure \ref{fig:dectree}), while shallower nodes show CATE estimates averaged within the subgroup defined by the corresponding split rule. The first split is on the variable ``Mother's level of education", specifically on whether the mother has attended college or not. The second level features a split on whether mother's ethnicity is white in one branch, and a split on whether mother has finished high school in the other. These are exactly the same covariates selected by the posterior splitting probabilities. The last set of splits is again on mother's ethnicity, number of children the mother has given birth to and whether child's birth weight is less than 2kg. Within these subgroups, CATE estimates range from a minimum of +2.1 to a maximum of +12.
	
	Both CATE's posterior splitting probabilities as well as subgroup analysis particularly point to covariates related to mother's education and ethnicity, in addition to birth weight (in the subgroup analysis only). Results concerning heterogeneity stemming from mother's ethnicity and child's birth weight are consistent with those in the original \citep{brooks_1992} and follow-up studies \cite{brooks_1994, brooks_1997}, where the treatment effect is found to be lower for white mothers and for children with lower weight. The advantage of carrying out subgroup analysis through models such as Shrinkage BCF lies in the fact that subgroup identification can be done ex-post using CATE estimates, without the need of manually identifying the groups or partitioning the original sample ex-ante.

	This illustrative example showed how Shrinkage BCF detects covariates which are responsible for the heterogeneity behind treatment impact in an example of real-world analysis, and how simple a posteriori partitioning of CATE estimates allows the derivation of optimal splitting rules to identify the most homogeneous subgroups in terms of treatment response. The analysis demonstrated that the estimation of individual (or subgroup) effects is a key factor for the correct evaluation and design of treatment administration policies.

	\section{Conclusions}  \label{sec:conclu}
	
	In this work, we introduced a sparsity-inducing version of the popular nonparametric regression model Bayesian Causal Forest, recently developed by \cite{hahn_2020}, in the context of heterogeneous treatment effects estimation. The new version proposed, Shrinkage Bayesian Causal Forest, is based on the contributions by \cite{linero_2018_sparse, linero_smooth}, and differs in the two additional priors that modify the way the model selects the covariates to split on. Shrinkage BCF allows targeted feature shrinkage on the prognostic score and CATE surfaces, and in addition returns posterior splitting probabilities, an interpretable measure of feature importance. In Section \ref{sec:simul}, we demonstrated its performance on simulated exercises that mimic confounded observational studies where only some covariates are relevant, while the rest of them constitutes nuisance predictors that can cause bias if included in a fully-saturated outcome model. Shrinkage BCF demonstrates competitive performance and scalability compared to the original version of BCF and to other state-of-the-art methods for CATE estimation, that tend to deteriorate with an increasing number of covariates. We also showed that it effectively tackles strong confounding from targeted selection, a property inherited from the BCF parametrization, and illustrated its use on a real-world study.
	
	In the simulated studies of Section 4 and 5, in addition to those in the supplementary material, we have investigated Shrinkage BCF's performance on different sparse DGPs. When we consider non-sparse DGPs instead, with few but all relevant covariates, default BCF might be the preferable option, even though Shrinkage BCF would not incur in much higher error.
	
	Besides the implementation of feature shrinkage per se, the additional advantage of Shrinkage BCF specification is that the pair of Dirichlet priors placed on the splitting probabilities can be tailored to incorporate subject-matter knowledge about the importance and impact of the covariates, separately for prognostic score and CATE. Embedding of prior information in a Bayesian fashion represents a way of avoiding a completely agnostic model, that nonetheless benefits from the excellent predictive properties of a nonparametric regression algorithm such as BART. Hence, the informative version of Shrinkage BCF can be useful in applied studies with limited sample size, where a priori knowledge is possessed and can be efficiently incorporated without losing the benefits of using a powerful non-linear model.
	
	Finally, as highlighted in different parts of the manuscript, we stress how the main advantage of DART (and consequently Shrinkage BCF) over BART (and BCF) is very much computational and improves performance in sparse DGPs settings exclusively. The MCMC convergence in DART is faster, as the model is urged to split more and more eagerly along the most predictive features. However, DART and Shrinkage BCF do not perform variable selection explicitly. An interesting future direction would be to augment DART priors to include predictor-specific inclusion parameters, to completely select out irrelevant predictors.
	
	\nocite{*}
	
	\bibliography{Refs}

\begin{thebibliography}{57}
\providecommand{\natexlab}[1]{#1}
\providecommand{\url}[1]{\texttt{#1}}
\expandafter\ifx\csname urlstyle\endcsname\relax
  \providecommand{\doi}[1]{doi: #1}\else
  \providecommand{\doi}{doi: \begingroup \urlstyle{rm}\Url}\fi

\bibitem[Alaa and van~der Schaar(2018)]{vanderschaar_2018}
A.~Alaa and M.~van~der Schaar.
\newblock Limits of estimating heterogeneous treatment effects: Guidelines for
  practical algorithm design.
\newblock In \emph{Proceedings of the 35th International Conference on Machine
  Learning}, volume~80, pages 129--138, 2018.

\bibitem[Alaa and van~der Schaar(2017)]{vanderschaar_2017}
A.~M. Alaa and M.~van~der Schaar.
\newblock Bayesian inference of individualized treatment effects using
  multi-task gaussian processes.
\newblock In \emph{Proceedings of the 31st International Conference on Neural
  Information Processing Systems}, NIPS'17, page 3427–3435, 2017.

\bibitem[Angrist and Pischke(2009)]{RePEc:pup:pbooks:8769}
J.~Angrist and J.-S. Pischke.
\newblock \emph{Mostly Harmless Econometrics: An Empiricist's Companion}.
\newblock Princeton University Press, 1 edition, 2009.

\bibitem[Angrist et~al.(1996)Angrist, Imbens, and Rubin]{angrist_1996}
J.~D. Angrist, G.~W. Imbens, and D.~B. Rubin.
\newblock Identification of causal effects using instrumental variables.
\newblock \emph{Journal of the American Statistical Association}, 91\penalty0
  (434):\penalty0 444--455, 1996.

\bibitem[Athey and Imbens(2016)]{athey_2016}
S.~Athey and G.~Imbens.
\newblock Recursive partitioning for heterogeneous causal effects.
\newblock \emph{Proceedings of the National Academy of Sciences}, 113\penalty0
  (27):\penalty0 7353--7360, 2016.

\bibitem[Breiman(2001)]{breiman_2001}
L.~Breiman.
\newblock Random forests.
\newblock \emph{Mach. Learn.}, 45\penalty0 (1):\penalty0 5–32, Oct. 2001.

\bibitem[Brooks-Gunn et~al.(1992)Brooks-Gunn, ruey Liaw, and
  Klebanov]{brooks_1992}
J.~Brooks-Gunn, F.~ruey Liaw, and P.~K. Klebanov.
\newblock Effects of early intervention on cognitive function of low birth
  weight preterm infants.
\newblock \emph{The Journal of Pediatrics}, 120\penalty0 (3):\penalty0
  350--359, 1992.

\bibitem[Brooks-Gunn et~al.(1994)Brooks-Gunn, McCarton, Casey, McCormick,
  Bauer, Bernbaum, Tyson, Swanson, Bennett, and Scott]{brooks_1994}
J.~Brooks-Gunn, C.~McCarton, P.~Casey, M.~McCormick, C.~Bauer, J.~Bernbaum,
  J.~Tyson, M.~Swanson, F.~Bennett, and D.~Scott.
\newblock Early intervention in low-birth-weight premature infants. results
  through age 5 years from the infant health and development program.
\newblock \emph{JAMA}, 272\penalty0 (16), October 1994.

\bibitem[Caron et~al.(2020)Caron, Manolopoulou, and Baio]{caron_2020}
A.~Caron, I.~Manolopoulou, and G.~Baio.
\newblock Estimating individual treatment effects using non-parametric
  regression models: a review.
\newblock \emph{arXiv:2009.06472}, 2020.

\bibitem[Chipman et~al.(1998)Chipman, George, and McCulloch]{cgm_1998}
H.~A. Chipman, E.~I. George, and R.~E. McCulloch.
\newblock Bayesian {CART} model search.
\newblock \emph{Journal of the American Statistical Association}, 93\penalty0
  (443):\penalty0 935--948, 1998.

\bibitem[Chipman et~al.(2010)Chipman, George, and McCulloch]{chipman_2010}
H.~A. Chipman, E.~I. George, and R.~E. McCulloch.
\newblock {BART}: Bayesian additive regression trees.
\newblock \emph{Ann. Appl. Stat.}, 4\penalty0 (1):\penalty0 266--298, 03 2010.

\bibitem[Dawid(2000)]{dawid_2000}
A.~P. Dawid.
\newblock Causal inference without counterfactuals.
\newblock \emph{Journal of the American Statistical Association}, 95\penalty0
  (450):\penalty0 407--424, 2000.

\bibitem[Dawid(2015)]{dawid_2014}
A.~P. Dawid.
\newblock Statistical causality from a decision-theoretic perspective.
\newblock \emph{Annual Review of Statistics and Its Application}, 2\penalty0
  (1):\penalty0 273--303, 2015.

\bibitem[Fan et~al.(2020)Fan, Hsu, Lieli, and Zhang]{fan_2020}
Q.~Fan, Y.-C. Hsu, R.~P. Lieli, and Y.~Zhang.
\newblock Estimation of conditional average treatment effects with
  high-dimensional data.
\newblock \emph{Journal of Business \& Economic Statistics}, 0\penalty0
  (0):\penalty0 1--15, 2020.

\bibitem[Foster et~al.(2011)Foster, m.~G.~Taylor, and Ruberg]{foster_2011}
J.~C. Foster, J.~m.~G.~Taylor, and S.~J. Ruberg.
\newblock Subgroup identification from randomized clinical trial data.
\newblock \emph{Statistics in medicine}, 30 24:\penalty0 2867--80, 2011.

\bibitem[Green and Kern(2012)]{green_2012}
D.~P. Green and H.~L. Kern.
\newblock {Modeling Heterogeneous Treatment Effects in Survey Experiments with
  Bayesian Additive Regression Trees}.
\newblock \emph{Public Opinion Quarterly}, 76\penalty0 (3):\penalty0 491--511,
  09 2012.

\bibitem[Hahn et~al.(2018)Hahn, Carvalho, Puelz, and He]{hahn_2018}
P.~R. Hahn, C.~M. Carvalho, D.~Puelz, and J.~He.
\newblock Regularization and confounding in linear regression for treatment
  effect estimation.
\newblock \emph{Bayesian Anal.}, 13\penalty0 (1):\penalty0 163--182, 03 2018.

\bibitem[Hahn et~al.(2020)Hahn, Murray, and Carvalho]{hahn_2020}
P.~R. Hahn, J.~S. Murray, and C.~M. Carvalho.
\newblock Bayesian regression tree models for causal inference: Regularization,
  confounding, and heterogeneous effects.
\newblock \emph{Bayesian Anal.}, 2020.

\bibitem[Hartford et~al.(2017)Hartford, Lewis, Leyton-Brown, and
  Taddy]{hartford_2017}
J.~Hartford, G.~Lewis, K.~Leyton-Brown, and M.~Taddy.
\newblock Deep {IV}: A flexible approach for counterfactual prediction.
\newblock In \emph{Proceedings of the 34th International Conference on Machine
  Learning}, volume~70, pages 1414--1423, 2017.

\bibitem[Hastie et~al.(2001)Hastie, Tibshirani, and Friedman]{tibshirani_2009}
T.~Hastie, R.~Tibshirani, and J.~Friedman.
\newblock \emph{The Elements of Statistical Learning}.
\newblock Springer Series in Statistics. Springer New York Inc., 2001.

\bibitem[Heckman(1979)]{heckman_1979}
J.~J. Heckman.
\newblock Sample selection bias as a specification error.
\newblock \emph{Econometrica}, 47\penalty0 (1):\penalty0 153--161, 1979.

\bibitem[Herren and Hahn(2020)]{herren_2020}
A.~Herren and P.~R. Hahn.
\newblock Semi-supervised learning and the question of true versus estimated
  propensity scores.
\newblock \emph{arXiv:2009.06183}, 2020.

\bibitem[Hill(2011)]{hill_2011}
J.~L. Hill.
\newblock Bayesian nonparametric modeling for causal inference.
\newblock \emph{Journal of Computational and Graphical Statistics}, 20\penalty0
  (1):\penalty0 217--240, 2011.

\bibitem[Holland(1986)]{holland_1986}
P.~W. Holland.
\newblock Statistics and causal inference.
\newblock \emph{Journal of the American Statistical Association}, 81\penalty0
  (396):\penalty0 945--960, 1986.

\bibitem[Imbens(2004)]{imbens_2004}
G.~W. Imbens.
\newblock Nonparametric estimation of average treatment effects under
  exogeneity: A review.
\newblock \emph{The Review of Economics and Statistics}, 86\penalty0
  (1):\penalty0 4--29, 2004.

\bibitem[Imbens and Rubin(2015)]{imbens_rubin_2015}
G.~W. Imbens and D.~B. Rubin.
\newblock \emph{Causal Inference for Statistics, Social, and Biomedical
  Sciences: An Introduction}.
\newblock Cambridge University Press, 2015.

\bibitem[Jacob et~al.(2017)Jacob, Murray, Holmes, and Robert]{jacob_2017}
P.~E. Jacob, L.~M. Murray, C.~C. Holmes, and C.~P. Robert.
\newblock Better together? statistical learning in models made of modules,
  2017.

\bibitem[Johansson et~al.(2016)Johansson, Shalit, and Sontag]{johansson_2016}
F.~D. Johansson, U.~Shalit, and D.~Sontag.
\newblock Learning representations for counterfactual inference.
\newblock In \emph{Proceedings of the 33rd International Conference on
  International Conference on Machine Learning - Volume 48}, page 3020–3029,
  2016.

\bibitem[King and Nielsen(2019)]{king_2018}
G.~King and R.~Nielsen.
\newblock Why propensity scores should not be used for matching.
\newblock \emph{Political Analysis}, 2019.

\bibitem[Künzel et~al.(2017)Künzel, Sekhon, Bickel, and Yu]{kunzel_2017}
S.~Künzel, J.~Sekhon, P.~Bickel, and B.~Yu.
\newblock Meta-learners for estimating heterogeneous treatment effects using
  machine learning.
\newblock \emph{Proceedings of the National Academy of Sciences}, 116, 06 2017.

\bibitem[Linero(2018)]{linero_2018_sparse}
A.~R. Linero.
\newblock Bayesian regression trees for high-dimensional prediction and
  variable selection.
\newblock \emph{Journal of the American Statistical Association}, 113\penalty0
  (522):\penalty0 626--636, 2018.

\bibitem[Linero and Yang(2018)]{linero_smooth}
A.~R. Linero and Y.~Yang.
\newblock Bayesian regression tree ensembles that adapt to smoothness and
  sparsity.
\newblock \emph{Journal of the Royal Statistical Society: Series B (Statistical
  Methodology)}, 80\penalty0 (5):\penalty0 1087--1110, 2018.

\bibitem[Lu et~al.(2018)Lu, Sadiq, Feaster, and Ishwaran]{lu_sadiq_2018}
M.~Lu, S.~Sadiq, D.~J. Feaster, and H.~Ishwaran.
\newblock Estimating individual treatment effect in observational data using
  random forest methods.
\newblock \emph{Journal of Computational and Graphical Statistics}, 27\penalty0
  (1):\penalty0 209--219, 2018.

\bibitem[McCarton et~al.(1997)McCarton, Brooks-Gunn, Wallace, Bauer, Bennett,
  Bernbaum, Broyles, Casey, McCormick, Scott, Tyson, Tonascia, and
  Meinert]{brooks_1997}
C.~McCarton, J.~Brooks-Gunn, I.~Wallace, C.~Bauer, F.~Bennett, J.~Bernbaum,
  R.~Broyles, P.~Casey, M.~McCormick, D.~Scott, J.~Tyson, J.~Tonascia, and
  C.~Meinert.
\newblock Results at age 8 years of early intervention for low-birth-weight
  premature infants. the infant health and development program.
\newblock \emph{JAMA}, 277\penalty0 (2):\penalty0 126—132, January 1997.

\bibitem[McCormick(1985)]{mccormick_1985}
M.~McCormick.
\newblock The contribution of low birth weight to infant mortality and
  childhood morbidity.
\newblock \emph{The New England journal of medicine}, 312\penalty0 (2), January
  1985.

\bibitem[McCormick et~al.(1990)McCormick, Gortmaker, and Sobol]{mccormick_1990}
M.~C. McCormick, S.~L. Gortmaker, and A.~M. Sobol.
\newblock Very low birth weight children: Behavior problems and school
  difficulty in a national sample.
\newblock \emph{The Journal of Pediatrics}, 117\penalty0 (5):\penalty0
  687--693, 1990.

\bibitem[Morris et~al.(2019)Morris, White, and Crowther]{morris_2019}
T.~P. Morris, I.~R. White, and M.~J. Crowther.
\newblock Using simulation studies to evaluate statistical methods.
\newblock \emph{Statistics in Medicine}, 38\penalty0 (11):\penalty0 2074--2102,
  2019.

\bibitem[Nie and Wager(2020)]{wager_2019}
X.~Nie and S.~Wager.
\newblock {Quasi-oracle estimation of heterogeneous treatment effects}.
\newblock \emph{Biometrika}, 09 2020.

\bibitem[Pearl(2009{\natexlab{a}})]{pearl_2009}
J.~Pearl.
\newblock \emph{Causality: Models, Reasoning and Inference}.
\newblock Cambridge University Press, USA, 2nd edition, 2009{\natexlab{a}}.

\bibitem[Pearl(2009{\natexlab{b}})]{pearl_letter_2009}
J.~Pearl.
\newblock Remarks on the method of propensity score.
\newblock \emph{Statistics in Medicine}, 28\penalty0 (9):\penalty0 1415--1416,
  2009{\natexlab{b}}.

\bibitem[Pearl(2018)]{Pearl_2018}
J.~Pearl.
\newblock Theoretical impediments to machine learning with seven sparks from
  the causal revolution.
\newblock \emph{Proceedings of the Eleventh ACM International Conference on Web
  Search and Data Mining - WSDM ’18}, 2018.

\bibitem[Powers et~al.(2018)Powers, Qian, Jung, Schuler, Shah, Hastie, and
  Tibshirani]{powers_2018}
S.~Powers, J.~Qian, K.~Jung, A.~Schuler, N.~H. Shah, T.~Hastie, and
  R.~Tibshirani.
\newblock Some methods for heterogeneous treatment effect estimation in high
  dimensions.
\newblock \emph{Statistics in Medicine}, 37\penalty0 (11):\penalty0 1767--1787,
  2018.

\bibitem[Robinson(1988)]{robinson_1988}
P.~M. Robinson.
\newblock Root-n-consistent semiparametric regression.
\newblock \emph{Econometrica}, 56\penalty0 (4):\penalty0 931--954, 1988.

\bibitem[Rosenbaum and Rubin(1983)]{rosenbaum_1983}
P.~R. Rosenbaum and D.~B. Rubin.
\newblock {The central role of the propensity score in observational studies
  for causal effects}.
\newblock \emph{Biometrika}, 70\penalty0 (1):\penalty0 41--55, 04 1983.

\bibitem[Ro\v{c}kov\'a and Saha(2019)]{rockova}
V.~Ro\v{c}kov\'a and E.~Saha.
\newblock On theory for {BART}.
\newblock In \emph{Proceedings of the Twenty-Second International Conference on
  Artificial Intelligence and Statistics}, volume~89, pages 2839--2848, 2019.

\bibitem[Rubin(1978)]{rubin1978}
D.~B. Rubin.
\newblock Bayesian inference for causal effects: The role of randomization.
\newblock \emph{Ann. Statist.}, 6\penalty0 (1):\penalty0 34--58, 01 1978.

\bibitem[Schuler et~al.(2018)Schuler, Baiocchi, Tibshirani, and
  Shah]{tibshirani_2018}
A.~Schuler, M.~Baiocchi, R.~Tibshirani, and N.~Shah.
\newblock A comparison of methods for model selection when estimating
  individual treatment effects, 2018.

\bibitem[Shalit et~al.(2017)Shalit, Johansson, and Sontag]{shalit_2017}
U.~Shalit, F.~D. Johansson, and D.~Sontag.
\newblock Estimating individual treatment effect: Generalization bounds and
  algorithms.
\newblock In \emph{Proceedings of the 34th International Conference on Machine
  Learning - Volume 70}, page 3076–3085, 2017.

\bibitem[Sivaganesan et~al.(2017)Sivaganesan, Müller, and Huang]{muller_2017}
S.~Sivaganesan, P.~Müller, and B.~Huang.
\newblock Subgroup finding via bayesian additive regression trees.
\newblock \emph{Statistics in Medicine}, 36\penalty0 (15):\penalty0 2391--2403,
  2017.

\bibitem[Starling et~al.(2019)Starling, Murray, Lohr, Aiken, Carvalho, and
  Scott]{tsbcf_2019}
J.~Starling, J.~Murray, P.~Lohr, A.~Aiken, C.~Carvalho, and J.~Scott.
\newblock Targeted smooth bayesian causal forests: An analysis of heterogeneous
  treatment effects for simultaneous versus interval medical abortion regimens
  over gestation, 05 2019.

\bibitem[Traskin and Small(2011)]{traskin_2011}
M.~Traskin and D.~S. Small.
\newblock Defining the study population for an observational study to ensure
  sufficient overlap: A tree approach.
\newblock \emph{Statistics in Biosciences}, 3:\penalty0 94--118, 2011.

\bibitem[Wager and Athey(2018)]{athey_2019}
S.~Wager and S.~Athey.
\newblock Estimation and inference of heterogeneous treatment effects using
  random forests.
\newblock \emph{Journal of the American Statistical Association}, 113\penalty0
  (523):\penalty0 1228--1242, 2018.

\bibitem[Yao et~al.(2018)Yao, Li, Li, Huai, Gao, and Zhang]{yao_2018}
L.~Yao, S.~Li, Y.~Li, M.~Huai, J.~Gao, and A.~Zhang.
\newblock Representation learning for treatment effect estimation from
  observational data.
\newblock In \emph{Advances in Neural Information Processing Systems 31}, pages
  2633--2643. 2018.

\bibitem[Zhao et~al.(2018)Zhao, Small, and Ertefaie]{zhao_2018}
Q.~Zhao, D.~S. Small, and A.~Ertefaie.
\newblock Selective inference for effect modification via the lasso.
\newblock \emph{arXiv:1705.08020}, 2018.

\bibitem[Zigler et~al.(2013)Zigler, Watts, Yeh, Wang, Coull, and
  Dominici]{zigler_2013}
C.~Zigler, K.~Watts, R.~Yeh, Y.~Wang, B.~Coull, and F.~Dominici.
\newblock Model feedback in bayesian propensity score estimation.
\newblock \emph{Biometrics}, 69, 02 2013.

\bibitem[Zigler and Dominici(2014)]{zigler_2014}
C.~M. Zigler and F.~Dominici.
\newblock Uncertainty in propensity score estimation: Bayesian methods for
  variable selection and model-averaged causal effects.
\newblock \emph{Journal of the American Statistical Association}, 109\penalty0
  (505):\penalty0 95--107, 2014.

\bibitem[Zimmert and Lechner(2019)]{zimmert_2019}
M.~Zimmert and M.~Lechner.
\newblock Nonparametric estimation of causal heterogeneity under
  high-dimensional confounding.
\newblock \emph{arXiv:1908.08779}, 2019.

\end{thebibliography}

\clearpage
	
	\appendix
	
\section{Additional Simulated Experiments}  \label{appenA}
	
	\subsection{Perfectly known propensity scores}
	
	Table \ref{tab:tarsel2} displays results obtained from Section 5.2 simulated exercize, where PS is assumed to be known and thus not estimated. Results are averaged over $H=250$ simulations.
	
	\begin{table}[H]
		\caption{\small $\text{Bias}$, $\sqrt{\text{PEHE}}$, 95\% Coverage and posterior splitting probability on the true $\pi(x_i)$ --- $(s_{\pi} \mid u_{\pi})$ --- for: i) default BCF; ii) Shrinkage BCF; iii) Shrinkage BCF without the true $\pi(x_i)$; iv) informative prior BCF with $k_{PS} = 50$; v) informative prior BCF with $k_{PS} = 100$.}
		\label{tab:tarsel2}
		\small
		\centering
		\begin{tabular}{l | ccc | c}
			\toprule
			\multicolumn{1}{c}{\textbf{Model}}  &  $\text{Bias}$  & $\sqrt{\text{PEHE}}$ & 95\% Coverage & $(s_{\pi} \mid u_{\pi})$ \\
			\midrule
			i) BCF  &  -0.03 $\pm$ 0.01  &  0.38 $\pm$ 0.02  &  0.95 $\pm$ 0.00 &  9.1\% \\
			ii) SH-BCF  & -0.02 $\pm$ 0.01  &  0.31 $\pm$ 0.02 & 0.97 $\pm$ 0.00 &  95.5\% \\
			iii) SH-BCF (no PS)   &  -0.06 $\pm$ 0.01  &  0.39  $\pm$  0.02  &  0.96 $\pm$ 0.01  &  - \\
			iv) I-BCF ($k_{PS} = 50$)    &  -0.02 $\pm$ 0.01  &  0.31 $\pm$ 0.02 & 0.97 $\pm$ 0.00 &  96.9\% \\
			v) I-BCF ($k_{PS} = 100$)    &  -0.02 $\pm$ 0.01  &  0.31 $\pm$ 0.02 & 0.97 $\pm$ 0.00 &  96.9\%  \\
			
			\bottomrule
		\end{tabular}
	\end{table}

	\subsection{Computational advantage of DART}
	
	In this small experiment, we briefly illustrate some of the computational advantages of DART's backfitting MCMC versus default BART. To this end, we compare on a purely predictive task three different specifications: i) default BART run for $6\,000$ MCMC draws (of which $4\,000$ burn-in); ii) long-chain BART run for $60,000$ MCMC draws ($40\,000$ burn-in); iii) DART run for $6\,000$ MCMC draws ($4\,000$ burn-in). The task is to predict $Y_i$ given $P=50$ predictors $\bm{X}_i$, of which only 5 are relevant, with $N=500$. The predictors $\bm{X}_i$ are simulated from a Gaussian Copula where elements of the correlation matrix are $\Theta_{jk} = 0.3^{|j - k|} + 0.1\mathbb{I}(j \neq k) $. Half of the predictors are continuous and half binary. The outcome $Y_i$ reads instead:
	$$ Y_i = 5 + 5 \sin(\pi X_{i,1}) + 2.5 (X_{i,2} - 0.5)^2 + 1.5 |X_{i,3}| + 2 X_{i,4} (X_{i,20} + 1) + \varepsilon_i \, ,$$
	where $\varepsilon_i \sim \mathcal{N} (0, 1)$. 
	
		\begin{table}[H]
		\caption{\small Test set $\text{RMSE}$ and average number of splits on the five relevant predictors, plus/minus 95\% Monte Carlo standard error for: i) default BART; ii) long-chain BART; iii) DART.}
		\label{tab:appen2}
		\small
		\centering
		\begin{tabular}{c | ccc }
			\toprule
			\multicolumn{1}{c}{\textbf{}}  &  BART & long BART & DART \\
			\midrule
			RMSE &  2.11 $\pm$ 0.03 & 1.99 $\pm$ 0.03 & 1.74 $\pm$ 0.03 \\
			\midrule
			$X_1$ &  21.85 $\pm$ 0.20 & 21.77 $\pm$ 0.18 & 66.43 $\pm$ 1.30 \\
			$X_2$ &  18.03 $\pm$ 0.14 & 18.06 $\pm$ 0.13 & 43.94 $\pm$ 0.72 \\
			$X_3$ &  6.85 $\pm$ 0.10 & 6.81 $\pm$ 0.09 & 10.49 $\pm$ 0.28 \\
			$X_4$ &  11.18 $\pm$ 0.10 & 11.09 $\pm$ 0.09 & 24.04 $\pm$ 0.44 \\
			$X_{20}$ &  6.63 $\pm$ 0.07 & 6.51 $\pm$ 0.05 & 39.44 $\pm$ 1.14 \\
			\bottomrule
		\end{tabular}
	\end{table}	
	
	The purpose of this exercise is to investigate whether the relative performance (measured with RMSE) of DART erodes with respect to BART run for dramatically long chain. Results displaying averaged RMSE for test set (we considered 70\%-30\% train-test split), in addition to the average number of splits on the 5 relevant predictors for each model are depicted in Table \ref{tab:appen2}. Results are averaged over $H=500$ Monte Carlo replications. We can see that running BART for way longer chains results in improved performance over short-chain BART. This is due to the fact that BART's MCMC concentrates very slowly, while DART allows for much faster posterior concentration.

	\subsection{High-dimensional $P$}
	
	In this third additional simulated experiment we study what happens to BCF and SH-BCF with an increasing number of predictors $P$. To this end, we consider setups with $P \in \{5, 10, 50, 100, 150\}$ respectively. Sample size is fixed at $N=250$, and we run $H=200$ Monte Carlo replications for each different $P$. The covariates $\bm{X}_i$ are simulated from a Gaussian Copula where elements of the correlation matrix are $\Theta_{jk} = 0.3^{|j - k|} + 0.1\mathbb{I}(j \neq k) $. The DGP is the following:
	\begin{align} 
	\mu (\bm{x}_i) = & ~ 3 + 1.5 \sin (\pi X_{i,1}) + 0.5 (X_{i,2} - 0.5)^2 + 1.5 ( 2 - | X_{i,3} |) + X_{i,4} (X_{i,\frac{P}{2}} + 1) ~ , \nonumber \\
	\tau (\bm{x}_i) = & ~ 0.1 + 1 |X_{i,1} - 1 | (X_{i,\frac{P}{2}} + 2) ~ , \nonumber \\
	\pi (\bm{x}_i) = & ~ \Phi \big( -0.5 + 0.2 X_{i, 1} + 0.1 X_{i, 2} + 0.4 X_{i,\frac{P}{2}} + \nu_i \big) ~ , \label{eq:appen3} \\
	Z_i \sim  & ~  \text{Bernoulli} \big( \pi (\bm{x}_i) \big) ~ ,  \nonumber \\
	Y_i =  & ~  \mu (\bm{x}_i) + \tau (\bm{x}_i) Z_i + \varepsilon_i ~ , \quad \text{where} \quad \varepsilon_i \sim \mathcal{N} (0, \sigma^2) ~ , \nonumber
	\end{align}
	where: $\Phi(\cdot)$ is the standard Normal c.d.f.; $\nu_i \sim \text{Uniform} (0, 0.1) $ is uniform noise; error standard deviation is set to $\sigma = 0.5 \, \hat{\sigma}_{\mu}$, where $\hat{\sigma}_{\mu}$ is the sample standard deviation of the simulated $\mu (\bm{x}_i)$.

\begin{figure}[t]
	    \centering
	    \includegraphics[scale=0.52]{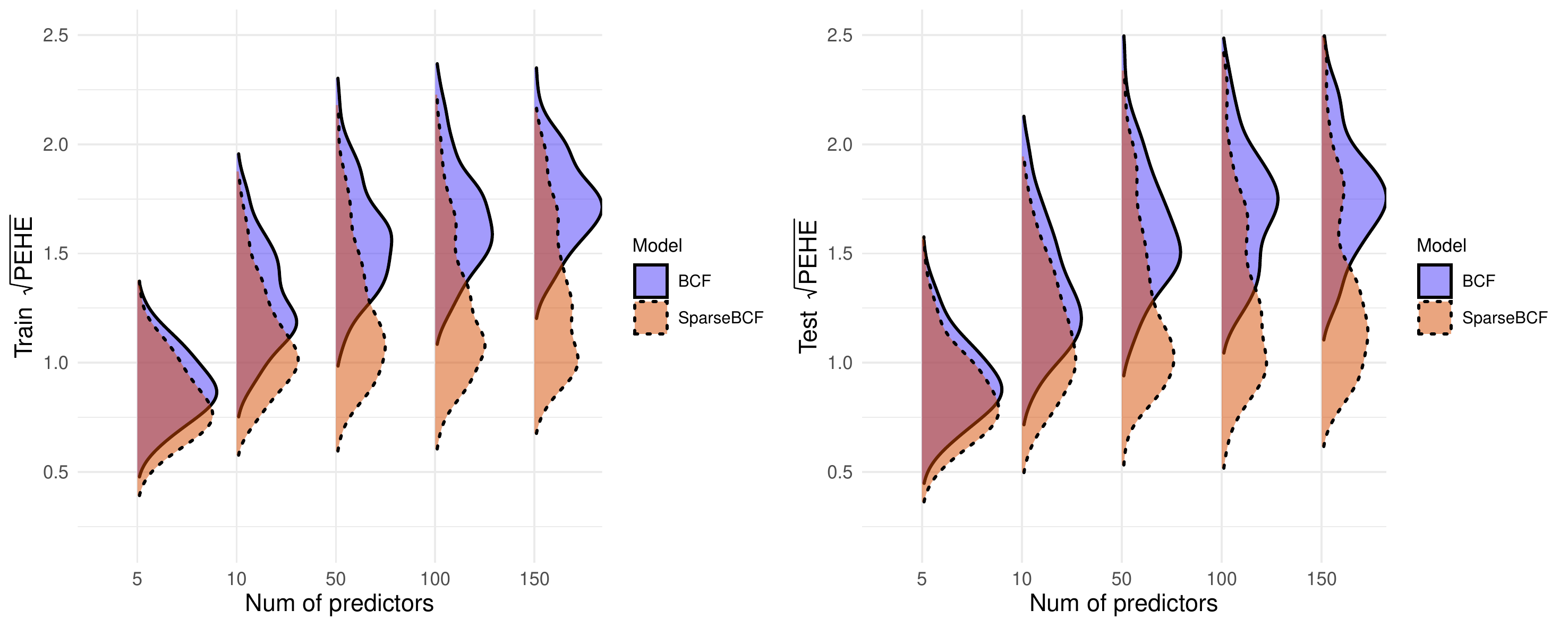}
	    \caption{Estimated train (left plot) and test (right plot) $\sqrt{\text{PEHE}}$ distributions generated by BCF and SH-BCF respectively, over an increasing number of predictors.}
	    \label{fig:my_label}
	\end{figure}

	\noindent A 70\%-30\% train-test set split is utilized. Results are shown in Table \ref{tab:appen3}, which depicts performance in terms of $\sqrt{\text{PEHE}}$, differentiated between train and test sets. We can appreciate how, compared to Sh-BCF, default BCF's performance deteriorates as $P$ increases, suffering from the curse of dimensionality.

\begin{table}[H]
		\caption{\small Train and test set average $\sqrt{\text{PEHE}}$, plus/minus 95\% Monte Carlo standard error, for BCF and SH-BCF with an increasing $P$.}
		\label{tab:appen3}
		\small
		\centering
		\begin{tabular}{c | cc | cc}
			\toprule
			     &  \multicolumn{2}{c|}{\textbf{BCF}} & \multicolumn{2}{c}{\textbf{SH-BCF}}  \\
			 \multicolumn{1}{c|}{$P$} & Train & Test & Train & Test \\
			 \midrule
			 5 & 0.91 $\pm$ 0.02 & 0.94 $\pm$ 0.03  & 0.83 $\pm$ 0.02 & 0.87 $\pm$ 0.03 \\
 			 10 & 1.30 $\pm$ 0.03 & 1.33 $\pm$ 0.04 & 1.12 $\pm$ 0.04 & 1.15 $\pm$ 0.04 \\
 			 50 & 1.57 $\pm$ 0.03 & 1.62 $\pm$ 0.04 & 1.23 $\pm$ 0.04 & 1.27 $\pm$ 0.05 \\
 			 100 & 1.66 $\pm$ 0.03 & 1.71 $\pm$ 0.04 & 1.26 $\pm$ 0.05 & 1.30 $\pm$ 0.05 \\
  			 150 & 1.74 $\pm$ 0.03 & 1.78 $\pm$ 0.04 & 1.32 $\pm$ 0.05 & 1.35 $\pm$ 0.06 \\
			\bottomrule
		\end{tabular}
	\end{table}

	\subsection{Different types of sparse DGPs}
	
	In the last extra simulated experiment, we study the performance of BCF and SH-BCF on four different types of sparse DGPs. In particular, we consider a setting with fixed $N=500$ and $P=5$, where covariates are generated again from a Gaussian Copula with correlation matrix elements set to $\Theta_{jk} = 0.3^{|j - k|} + 0.1\mathbb{I}(j \neq k) $. We run BCF and SH-BCF for $H=200$ Monte Carlo replications on each of the following four different versions of a DGP, according to what surface is generated as sparse:
	\begin{itemize}
	    \item[1)] \textbf{No Sparsity}. The first version features no sparsity at all, meaning that all the covariates are relevant for estimating every function of interest. The DGP reads:
	    \begin{align} 
    	\mu (\bm{x}_i) = & ~ 3 + 1.5 \sin (\pi X_{i,1}) + 0.5 (X_{i,2} - 0.5)^2 + 1.5 ( 2 - | X_{i,3} |) + 1.5 X_{i,4} (X_{i,5} + 1) ~ , \nonumber \\
    	\tau (\bm{x}_i) = & ~ 0.1 + 1 |X_{i,1} - 1 | (X_{i,5} + 2) - 0.4 X_{i,3} + 0.6 X_{i,2} X_{i,4} ~ , \nonumber \\
    	\pi (\bm{x}_i) = & ~ \Phi \big( -0.2 + 0.8 X_{i, 1} - 0.1 X_{i, 2} + 0.1 X_{i,3} X_{i,4} - 0.4 X_{i,5} + \nu_i \big) ~ , \label{eq:appen4} \\
    	Z_i \sim  & ~  \text{Bernoulli} \big( \pi (\bm{x}_i) \big) ~ ,  \nonumber \\
    	Y_i =  & ~  \mu (\bm{x}_i) + \tau (\bm{x}_i) Z_i + \varepsilon_i ~ , \quad \text{where} \quad \varepsilon_i \sim \mathcal{N} (0, 1) ~ , \nonumber
    	\end{align}
    	where $\Phi(\cdot)$ is the standard Normal c.d.f. and $\nu_i \sim \text{Uniform} (0, 0.1) $ is uniform noise
    	
    	\item[2)] \textbf{Sparse $\bm{\pi(\cdot)}$}. The second DGP is exactly the same as (\ref{eq:appen4}), but propensity score is a sparse surface, equal to $\pi (\bm{x}_i) = \Phi ( -0.2 + 0.8 X_{i, 1} + \nu_i )$
    	
    	\item[3)] \textbf{Sparse $\bm{\mu (\bm{x}_i)}$}. The third DGP is the same as (\ref{eq:appen4}), but prognostic score is a sparse surface, equal to $\mu (\bm{x}_i) = 3 + 1.5 ( 2 - | X_{i,3} |)$
    	
    	\item[4)] \textbf{Sparse $\bm{\tau (\bm{x}_i)}$}. Finally, the last DGP is the same as (\ref{eq:appen4}), but treatement effect is a sparse surface $\tau (\bm{x}_i) = 0.1 + 1 |X_{i,1} - 1 |$
	\end{itemize}
	
	\noindent Again, a 70\%-30\% train-test split is considered. Table \ref{tab:appen4} shows BCF's and SH-BCF's results in terms of train and test $\sqrt{\text{PEHE}}$.
	
		\begin{table}[H]
		\caption{\small Train and test set $\sqrt{\text{PEHE}}$, plus/minus 95\% Monte Carlo standard error, for BCF ans SH-BCF on the four different version of sparse DGPs.}
		\label{tab:appen4}
		\small
		\centering
		\begin{tabular}{l | cc | cc}
			\toprule
			     &  \multicolumn{2}{c|}{\textbf{BCF}} & \multicolumn{2}{c}{\textbf{SH-BCF}}  \\
			 \multicolumn{1}{c|}{DGP} & Train & Test & Train & Test \\
			 \midrule
			 Not Sparse & 0.96 $\pm$ 0.02 & 0.98 $\pm$ 0.03 & 0.92 $\pm$ 0.02 & 0.95 $\pm$ 0.03 \\
 			 Sparse $\pi(\cdot)$ & 0.91 $\pm$ 0.02 & 0.93 $\pm$ 0.03 & 0.88 $\pm$ 0.02 & 0.93 $\pm$ 0.03  \\
 			 Sparse $\mu(\cdot)$ & 0.85 $\pm$ 0.02 & 0.87 $\pm$ 0.03 & 0.81 $\pm$ 0.02 & 0.83 $\pm$ 0.03 \\
 			 Sparse $\tau(\cdot)$ & 0.67 $\pm$ 0.02 & 0.67 $\pm$ 0.02  & 0.66 $\pm$ 0.02 & 0.66 $\pm$ 0.02  \\
			\bottomrule
		\end{tabular}
	\end{table}

	\section{Variables included in the analysis}  \label{appenB}
	
	Table \ref{tab:vars} here below provide a full list of variables used for the analysis in Section 6.
	
	\begin{longtable}{l|l|l}
		\caption{Variables from the Infant Health and Development Program (IHDP)} \\
		\centering \footnotesize
		\textbf{Variable}  &  \textbf{Description} & \textbf{Type} \\
		\midrule
		\textit{iq}  &  Score in IQ test (outcome $Y$)  &  Numeric 	\\	
		\textit{treat} & Participation to the program (treatment $Z$)  &	 Binary	\\
		\midrule
		\textit{bw} &  Child's weight at birth (in grams) &	 Numeric	\\
		\textit{momage} &  Mother's age  &	 Numeric	\\
		\textit{nnhealth} &  Neo-natal health index  &	 Numeric	\\
		\textit{birth.o} &  Child's order of birth  &  Numeric	\\
		\textit{parity} &  Number of children  &  Numery	\\
		\textit{moreprem} &  Number of children born prematurely & Numeric	\\
		\textit{cigs} &  Smoke during pregnancy  & Numeric	\\
		\textit{alcohol} &  Drinks during pregnancy  & Numeric	\\
		\textit{ppvt.imp} &  Mother's PPVT test result 1 year post birth  & Numeric	\\
		\textit{bw\_2000} &  Birth weight above/below 2kg  & Binary	\\
		\textit{female} &  Child is a female  & Binary	\\
		\textit{mlt.birt} &  Number of multiple births  & Ordinal	\\
		\textit{b.marry} &  Marital status at birth  &  Binary	\\
		\textit{livwho} &  What family member lives with the child  &  Ordinal	\\
		
		\textit{language} &  Language spoken at home &  Binary	\\
		\textit{whenpren} &  Trimester when prenatal care started  &  Ordinal	\\
		\textit{drugs} &  Drug use during pregnancy  &  Binary	\\
		\textit{othstudy} &  Participating in other studies at the same time  &  Binary	\\
		\textit{site1} &  Site number 1 &  Binary	\\
		\vdots &  \vdots  &  \vdots	\\
		\textit{site8} &  Site number 8  &  Binary	\\
		\textit{momblack} &  Mother's ethnicity black  &  Binary	\\
		\textit{momhisp} &  Mother's ethnicity hispanic  &  Binary	\\
		\textit{momwhite} &  Mother's ethnicity white  &  Binary	\\
		\textit{workdur.imp} &  Mother worked during pregnancy  &  Binary	\\
		\textit{momed4F} &  Mother's education level  &  Ordinal
		
		\label{tab:vars}
	\end{longtable}

\end{document}